\documentclass[12pt,russian,english,dvips,cite]{extarticle}
\usepackage[T1]{fontenc}
\usepackage[koi8-r,latin9]{inputenc}
\usepackage{array}
\usepackage{rotfloat}
\usepackage{multirow}
\usepackage{amsmath}
\usepackage{amssymb}
\usepackage{graphicx}
\usepackage{esint}

\makeatletter

\providecommand{\tabularnewline}{\\}

\newcommand{\lyxaddress}[1]{
\par {\raggedright #1
\vspace{1.4em}
\noindent\par}
}

\usepackage{babel}

\usepackage{babel}

\usepackage{babel}

\makeatother

\usepackage{babel}
\begin{document}

\title{Effective Hamiltonian Crystal Field for Magnetic Interactions in
Polynuclear Transition Metal Complexes. Sequential Derivation and
Exemplary Numerical Estimates.}

\author{Andrei L. Tchougréeff }

\maketitle

\lyxaddress{Moscow Center for Continuous Mathematical Education, Bol. Vlasevskiy
per., 14, Moscow 119002, Russia}
\begin{abstract}
By this we extend our work of the year 1992 devoted to calculating
the intrashell excitations in the \emph{d}-shells of coordination
compounds of the first transition metal row, which resulted in the
Effective Hamiltonian Crystal Field (EHCF) method, to their polynuclear
analogs in order to assure the description of several open \emph{d}-shells
and of magnetic interactions of the effective spins residing in these
shells. This is a challenging task since it requires improving the
precision of \emph{ca.} 1000 $\mathrm{cm}{}^{-1}$ (that of describing
the excitation energies of the single \emph{d}-shells by the already
well successful EHCF method) to the that of \emph{ca.} 10 $\div$
100 $\mathrm{cm}{}^{-1}$ characteristic for the energies required
to reorient the spins \emph{i.e.} eventually by two orders of magnitude.
This is performed within the same paradigm as used for the EHCF method:
the concerted usage of the McWeeny's group-function approximation
and the Löwdin partition technique. These are applied to develop the
effective description of the \emph{d}-system of the polynuclear complexes,
composed of several \emph{d}-shells, including the working formulae
for the exchange parameters between the \emph{d}-shells belonging
to different transition metal ions. These formulae are implemented
in the package MagAîxTic and tested against a series of binuclear
complexes of trivalent Cr and Fe cations featuring $\mu$-oxygen superexchange
paths in order to confirm the reproducibility of the trends in the
series of values of exchange parameters for the compounds differing
by the chemical substitutions and other details of composition and
structure. The results of calculations are in a reasonable agreement
with available experimental data and other theoretical methods. 
\end{abstract}

\section{Introduction}

Magnetic properties of transition metal complexes (TMC) derive from
the classical work reducing them to those of the respective isolated
\emph{d}-shells.\cite{Bethe} Those of so called polynuclear TMC (PTMC
- those containig two or more transition metal ions - TMI's - with
open \emph{d}-shells) in the low-energy range are usually as well
classically mapped to the effective exchange (spin) Dirac-Van Vleck-Heisenberg
Hamiltonian:\cite{Van-Vleck} 
\begin{equation}
\hat{H}_{\mathrm{DvVH}}=\sum_{i<j}J_{ij}\hat{S}_{i}\hat{S}_{j},\label{eq:DvVH}
\end{equation}
 controlled by the effective exchange parameters $J_{ij}$ describing
the interactions of the effective spins $\hat{S}_{i}$ residing in
the \emph{d}-shells of the corresponding TMI's with those in their
neighbourhood. Numerous attempts have been undertaken to derive reliable
numerical values of these parameters suitable to describe the lowest-energy
range of excitations of the electronic subsystem of PTMCs quantum
chemically. The complexity of this task (examples of the tentative
application objects from the recent literature\cite{Milios,Carretta,Mueller,Kortz}
are quite impressive) prevented, however, from the decisive success.
The theoretical considerations of the effective magnetic parameters
present in the literature are numerous, but can be classified in three
types: (i) \textit{ab initio} approaches, (ii) DFT-based approaches,
and (iii) semiempirical/semiquantitative approaches predominantly
stemming from the solid state physics realm.

The \textit{ab initio} methods are potentially capable to solve the
problem (see below).\cite{DeLothCassouxDaudeyMalrieu,CharlotVerdaguerJournauxDeLothDaudey,deLothDaudeyAstheimerWalzHaase,FinkFinkStaemmler,WangFinkStaemmler}
However, the systems of real interest in this area may contain thousands
of atoms of which hundreds are going to be transition metal ones.\cite{Milios,Carretta,Mueller,Kortz}
Within the \textit{ab initio} realm such systems require the account
of correlations (both dynamical and noindynamical) at a level which
causes the scaling of the requred computational resources as $N^{7}$
where $N$ is the number of orbitals involved. It will be hardly possible
to apply such an approach within a chemical problem setting that of
establishing/reproducing the trends in a series of similar compounds
rather obtaining a unique number for a unique system. The dependence
of the result on the basis state used has been established as well.\cite{deLothDaudeyAstheimerWalzHaase}

The DFT based methods recently received considerable attention in
this context. \cite{Ruiz1999,Ruiz2004} The success reported for the
DFT-based techniques\cite{Ruiz1999,Ruiz2004} heavily relies upon
the possibility to obtain the broken symmetry solutions in the UHF
setting for the respective molecules. Although trivial for binuclear
complexes, usual guinea pigs for the workers in the field, it may
become very problematic for the systems where odd-numbered cycles
of spins coupled through antiferromagnetic exchange interactions can
be singled out and thus one can expect frustration, so that the spins
cannot arrange themseves in a single strucutre describable by the
Ising model which is a prerequisite for applying either the DFT-based
technology or similar approach based on usage of nonempirical Hamiltonians.
\cite{Ruiz1999,Ruiz2004} %
\footnote{It must be understood, as well, that the above examples of experimentally
interesting systems \cite{Milios,Mueller,Carretta,Kortz} are still
not accessible for the DFT based methods either. Also one has to admit
that usage of the DFT methods for this purpose is problematic in a
more general sence: already reproducing the ground state total spin,
one-center \emph{d}-\emph{d} spectra and their geometry dependence
of the mononuclear TMCs represents a problem for them. Apparently
the precision requirements for estimating the magnetic interactions
in the polynuclear TMCs are much more tough than for the ground state
spin in mononuclear TMCs. Thus the DFT-based methods under certain
conditions can be tuned to reproduce the observable exchange parameters,
but the uniform picture of (P)TMCs evades the DFT methods.\cite{Tch074,Tch075} %
}

These complications made us to reconsider the ingenious idea due to
P.W. Anderson \cite{Anderson1959,Anderson1963} who suggested a two
step procedure eventually yielding the reliable estimate of the magnetic
exchange parameters without invoking \emph{a priori} unknown magnetic
ground states of a complex or a solid. According to the Anderson's
prescription (see also\cite{Goodenough,Vonsovsky}) in a line with
the original implicit picture used by Bethe\cite{Bethe} formulated
explicitly only later.\cite{Harrison} At the first step one has to
obtain one-electronic wave functions of the magnetic ions (in our
case these are the states of the TMI's \emph{d}-shells) immerced in
the insulating diamagnetic medium (ligands) within some kind the ligand
field theory.\cite{Ballhausen,Bersukerbook} At the second step one
has to estimate the interactions between electrons residing in the
magnetic orbitals defined at the first step.%
\footnote{Calculating system <<without magnetism>> in a solid state \emph{i.e.}
translationally invariant rather a <<crystal field>> setting would
lead to a metallc state with poorly definable Wannier states. That
is why one has to rely upon as well not well defined cluster models
for these ions. %
}

This way of thinking was pursued in our work\cite{Tch017} where we
performed the first step: the consistent description of the magnetic
states of the isolated TMIs. It is commonly believed that semi-empirical
methods of quantum chemistry are not of much use for TMCs. The reason
of such attitude originates from very narrow understanding of the
nature of semiempirical methods. Of course, if one takes them naïvely:
as a parameterized SCF in the valence basis, one cannot count on considerable
success. Indeed, such an approach is known to be unsuccessful for
decades in coping with the TMCs ground state total spins and with
their geometry dependence not talking about the low-energy $d$-$d$
excitation spectra. The way out had been found\cite{Tch017} in the
concerted usage of the McWeeny group-function formalism \cite{McWeenybook}
and of the Löwdin partition technique\cite{Lowdin} as described in
details in reviews\cite{Tch049,Tch081} and in monograph.\cite{Tch080}
This can be considered as a new concept of semiempirism in quantum
chemistry: taking the \emph{form} of the wave function on the basis
of observable electronic groups (like \emph{d}-shells, $\pi$-systems,
two-center bonds \emph{etc.}) characteristic for the considered class
of molecules. Taking this way allowed us previously to reach considerable
success in describing the ground states and $d$-$d$ excitation spectra
of TMCs\cite{Tch017,Tch029,Tch030,Tch033,Tch037} and of their geometry
dependence.\cite{Tch074,Tch075,Tch043,Tch044,Tch046,Tch060,Tch065,Tch070}
In the present work we concentrate on the derivation of the effective
Hamiltonian for the $d$-system of a PTMC which eventually contains
necessary effective magnetic terms and by this follow the way (iii)
of developing a substantiated semiempirical approach to estimating
the parameters of effective magnetic interactions suitable for sizeable
systems without addressing in advance their magnetic structure. This
way would not be possible without works\cite{WeiheGuedel1996,WeiheGuedel1997,WeiheGuedel1998,WeihGuedel2000,WeiheGuedelToftlund,TuczekSolomon,BrownRemarMusselmanSolomon}
where the quantitative empirical relations between the parameters
of the crystal field felt by the \emph{d}-shells in series of PTMCs
and their respective magnetic exchange parameters has been firmly
established. These phenomenological works allowed us to hope (and
as shown below with good grounds) that the estimates of the crystal
field parameters by semiempirical procedures assembled in the EHCF
method\cite{Tch017} apply as well to the magnetic exchange parameters
in PTMCs.

The paper is organized as follows. In the next Section we describe
the necessary modification of the EHCF formalism for the case of polynuclear
TMCs (PTMCs) and specific contributions to the effective Hamiltonian
of their \emph{d}-electrons. The details of the derivation are quite
cumbersome and thus transferred to the Appendices. In further two
Sections we describe respectively some relevant details of implementation
and results of our calculations as compared to previous ones and the
experiment. After that we discuss our results and the relation or
our approach with others existing in this area.

\section{Effective Hamiltonian of Crystal Field as modified for magnetic interactions}

The intuitive physical picture we are going to pursue is very simple.
The low-energy range of excitations of the electronic system of a
polynuclear transition metal complex (PTMC) is spanned by the states
differring by the relative orientations of the local spins (electronic
in origin) residing in the respective \emph{d}-shells. From the \emph{ab
initio} or DFT viewpoints such a picture is an enormous oversimplification
since everyone is aware of the fact that all the basis electronic
states are strongly hybridized. Nevertheless, the relative magnitudes
of the electron transfer (hopping) matrix elements in comparison to
the relevant energies of the one-electron charge transfer states differ
drastically for different states within the PTMC. Namely, this ratio
is by an order of magnitude larger for the transfers involving \emph{d}-AOs
than for a similar ratio for the transfers involving the AOs on the
close ligand atoms. This is the physical reason why in most cases
one can unequivocally ascribe some definite integer number of electrons
to the \emph{d}-shells and interpret the low-energy optical spectra
of (P)TMCs in terms of the states of such shells with fixed integer
numbers of electrons \emph{i.e.} with use of the crystal field theory.\cite{Ballhausen,Bersukerbook,Liver}
These ideas allowed us to make a formal move and to identify the parts
of systems according to their respective different physical regimes:
the highly correlated $d$-AOs which we term to span the one-electron
space of the \emph{d}-system and the weakly correlated AOs of the
<<organogenic>> ligand atoms and the weakly correlated \emph{sp}-AOs
of TMI's which span the one-electron space of the \emph{l}-system.
The difference of the physical conditions in so defined systems allows
us treat them differently as well. The $l$-system being in a low
correlated regime can be treated with an acceptable precision by a
semiempirical SCF theory (INDO will suffice, see below); the \emph{d}-system
requires in principle kind of correlated treatment. In order to be
able to treat so defined subsystems separately one has to develop
(mutually dependent) effective Hamiltonians for the singled out subsystems.
That for the \emph{d}-system eventually contains the required <<magnetic>>
terms. This development is sketched in the subsequent Subsections.

\subsection{Molecular Hamiltonian breakdown }

The basis of one-electron states for describing a PTMC is defined
within a general semiempirical scheme and is thus spanned by the valence
AOs which are considered to be implicitly orthogonal. For a PTMC the
valence AO basis is formed by the valence AOs of the ligand atoms
(largely, but not mandatory, organogenic) and by those of the TMIs
it contains. These latter are the 4\emph{s}-, 4\emph{p}-, and 3\emph{d}-AOs
for the atoms of the first transition series. On the basis of the
ratios of the characteristic intersubsystem hopping and Coulomb integrals
we separate as explained previously the entire set of the AOs into
the basis subsets spanning respectively \emph{d}- and \emph{l}-systems
of a PTMC. The corresponding orbital subspaces are thus implicitly
orthogonal.

According to the above classification of the valence AO's the semiempirical
Hamiltonian for a PTMC rewrites as (the explicit form of all operators
is given in Appendix \ref{sec:Hamiltonian-contributions}): 
\begin{equation}
\begin{array}{c}
H=H_{d}+H_{l}+H_{\mathrm{int}}\\
H_{\mathrm{int}}=H_{c}+H_{r}+H_{x}
\end{array}\label{eq:Hamiltonian}
\end{equation}
 where the operators $H_{d}$ (one for $d$-electrons of all TMIs
in the complex) and $H_{l}$ (that for electrons in the \emph{l}-system)
are formally defined as those involving the Fermi creation and annihilation
operators\cite{Surjanbook} referring only to the \emph{d}- and \emph{l}-systems,
respectively, whereas the interaction operator $H_{\mathrm{int}}$
involves mixed products of the Fermi creation and annihilation operators
referring to the \emph{d}- and \emph{l}-systems (see Appendix \ref{sub:Subsystem's-Hamiltonians}).

The interaction operator $H_{\mathrm{int}}$ is a sum of contributions,
which as well can be formally defined through the types of the products
of the Fermi operators entering them. Correspondingly, the operators
$H_{c}$ and $H_{x}$ are two-electron operators which involve the
products of four Fermi operators and represent, respectively, the
Coulomb and exchange interactions between two subsystems; the operator
$H_{r}$ is the one-electron operator which involves only products
of one Fermi operator referring to the \emph{d}- and one Fermi operator
referring to the \emph{l}-AO subspaces (details see in Appendix \ref{sub:Intersubsystem-interaction-opera}).

\subsection{Configuration subspaces and Löwdin partitions}

The main complication for a sequential description of magnetic properties
of PTMCs is the enormous number of electronic configurations to be
taken into account. Yet a simpler task of describing mononuclear TMCs
involved quite a number of intermediate (<<outer space>>) configurations
which were addressed with use of the Löwdin partition technique.\cite{Lowdin}
Its success as well as technical details of its implementation in
the EHCF method makes us to try similar moves as applied to PTMCs.
The Löwdin partition can be regarded as a specific method of separating
variables in a many-particle system such that upon applying it, the
system turns to be representable by a product function whose multipliers
describe otherwise \emph{entangled} states of the parts of the whole
system. That is to say that performing a Löwdin partition is a prerequisite
of employing the McWeeny group-function approximation. Thus our derivation
evolves as a sequence of the Löwdin partitions allowing to single
out relevant (<<model>>) configuration subspaces and to take into
account the effect of the <<outer space>> configurations outside
the model subspaces considered explicitly at each step on the matrix
elements of the model Hamiptonian.

\subsubsection{Separation of \emph{d}- and \emph{l}-electronic variables }

The first move in the EHCF derivation was the separation of the variables
in the single \emph{d}-shell available in a mononuclear TMC and in
the ligands. That finally allowed us to write the TMCs electronic
wave function as an antisymmetrized product of the group functions
for two distinguishable subsystems in the TMC, having physically different
conditions, the \emph{d}-shell and the rest: 
\begin{equation}
\Psi_{\mathrm{TMC}}=\Phi_{d}\wedge\Phi_{L},\label{eq:WaveFunction}
\end{equation}
 the form implicitly assumed by the crystal field theory.\cite{Bethe,Ballhausen,Bersukerbook}%
\footnote{Here $\wedge$ stands for the antisymmetrized product of the electronic
wave functions, if the Fermi operators are used this notation means
that all of them referring to the \emph{d}-shell stand to the left
of those referring to the ligand orbitals%
} The form eq. (\ref{eq:WaveFunction}) is approximate since the operator
$H_{r}$ acts and the electrons move between the subsystems so that
the entire TMC occurs in an electronic state where those of the individual
subsystems are \emph{entangled}. In order to recover the crystal field
theory picture in which the number of electrons in the \emph{d}-system
of the TMC is constant for the low-energy part of the electronic spectrum
the exact Hamiltonian has to be projected to the model subspace spanned
by the configurations in which the number of electrons in the $d$-shell(s)
is fixed. The same move has to be done for the PTMCs. Let the overall
number $n_{d}$ of electrons in the $d$-system be fixed. The numbers
of electrons in the individual $d$-shells satisfy the following conditions:
\begin{equation}
\begin{array}{ccc}
n_{l} & = & N-n_{d},\\
n_{d} & = & \sum_{i}n_{d}(i),
\end{array}\label{eq:NumberOfElectrons}
\end{equation}
 where $n_{d}(i)$ is the number of electrons in the \emph{d}-shell
of the $i$-th TMI and $N$ is the total number of electrons in the
PTMC. Since the states with different distributions of electrons among
the $d$-shells of the different TMIs enter the consideration (see
below) it may be technically easier to trace that $n_{l}$ -- the
number of electrons in the $l$-system -- remains constant. As in
the derivation of the EHCF method\cite{Tch017} the configurations
with numbers of $d$- (or \emph{l}-) electrons different from that
fixed above (ligand to metal and metal to ligand charge transfer states,
respectively abbreviated as LMCT and MLCT) have to be taken into account
to keep track of the effect of the resonance operator $H_{r}$.

As in the derivation of the EHCF method\cite{Tch017} it is done with
use the Löwdin partition technique.\cite{Lowdin} Let $P$ be the
projection operator to the model many-electron configuration subspace
spanned by the Slater determinants with $n_{l}$ electrons in the
\emph{l}-system and remaining $n_{d}$ in all the $d$-shells and
let $Q=1-P$ be its complementary projection operator. The Löwdin
partition replaces the original Hamiltonian acting in the entire configuration
space by the effective Hamiltonian $H^{\mathrm{eff}}(E)$ which acts
in the model configuration subspace $\mathrm{Im}P$: 
\begin{equation}
\begin{array}{c}
H^{\mathrm{eff}}(E)=PHP+PH_{rr}P,\\
H_{rr}=H_{r}QR(E)QH_{r},
\end{array}\label{eq:EffectiveHamiltonian}
\end{equation}
 but by construction has the same eigenvalues as the original Hamiltonian.
In eq. (\ref{eq:EffectiveHamiltonian}) the resolvent: 
\begin{equation}
R(E)\equiv(EQ-QHQ)^{-1}\label{eq:ResolventItalic}
\end{equation}
 is used. The form and the notation $H_{rr}$ is due to the fact that
the operator $H_{r}$, the only term which transfers electrons between
the $d$- and $l$-systems of the PTMC and by this mixes the states
in the $\mathrm{Im}P$ and $\mathrm{Im}Q$ configuration subspaces,
\begin{equation}
\begin{array}{c}
PHP=PH_{0}P+PH_{x}P\\
H_{0}=H_{d}+H_{l}+H_{c}\\
PHQ=PH_{r}Q;QHP=QH_{r}P\\
QHQ=QH_{0}Q+QH_{r}Q+QH_{x}Q
\end{array}\label{eq:ProjectorAction}
\end{equation}
 ultimately enters the answer in an even degree. So we arrive to the
effective Hamiltonian acting in the model configuration subspace with
the fixed number of electrons in the \emph{d}- and \emph{l}-systems
of a PTMC, but taking into account the LMCT and MLCT configurations
in the <<outer>> configuration subspace.

The resolvent eq. (\ref{eq:ResolventItalic}) can be rewritten as
a series with use of the bare resolvent: 
\begin{equation}
R_{0}(E)\equiv(EQ-QH_{0}Q)^{-1}\label{eq:BareResolvent}
\end{equation}
 derived from the operator $QH_{0}Q$ (conserving the quantities $n_{d}$
and $n_{l}$ which are thus <<good>> quantum numbers). The first
terms in the series yield: 
\begin{equation}
\begin{array}{c}
R(E)=(EQ-QH_{0}Q-QH_{r}Q-QH_{x}Q)^{-1}\approx\\
\approx R_{0}(E)+R_{0}(E)\left(QH_{r}Q+QH_{x}Q\right)R_{0}(E)+\\
+R_{0}(E)\left(QH_{r}Q+QH_{x}Q\right)R_{0}(E)\left(QH_{r}Q+QH_{x}Q\right)R_{0}(E)
\end{array}\label{eq:ResolventExpansion}
\end{equation}
 When inserted in eq. (\ref{eq:EffectiveHamiltonian}) the above expression
corresponds to the fourth order of perturbation theory with respect
to the resonance operator $H_{r}$ and at least second order with
respect to $H_{x}$. It is shown in Appendix \ref{sec:Expansion-of-the}
that the operator $H_{r}$ enters the effective Hamiltonian eq. (\ref{eq:EffectiveHamiltonian})
in a twofold way. First, the $H_{rr}^{(0)}$ component of $H_{rr}$
which results in no ionization of the \emph{l}-system (formal definition
of the compoments $H_{rr}^{(m\pm)}$ see Appendix \ref{sec:Expansion-of-the})
acts within the model subspace $\mathrm{Im}P$ of our interest. Second,
the $H_{rr}^{(2\pm)}$ components of $H_{rr}$ take into account the
<<doubly ionic>> MLCT and LMCT configurations in which the \emph{l}-system
is doubly ionized in either sense. We denote these configurations
as (LM)$^{2}$CT or (ML)$^{2}$CT with respect to the direction of
the charge transfer. These configurations have been shown to be important
for estimating the amount of effective antiferromagnetic interaction.\cite{WeiheGuedelToftlund}
Collecting the relevant terms yields:

\begin{eqnarray}
H^{\mathrm{eff}}(E) & \approx & PH_{0}P+PH_{x}P+PH_{rr}^{(0)}P+\nonumber \\
 & + & PH_{r}^{(+)}Q^{(1+)}R_{0}^{(1+)}Q^{(1+)}H_{x}Q^{(1+)}R_{0}^{(1+)}Q^{(1+)}H_{r}^{(+)}P+\nonumber \\
 & + & PH_{r}^{(-)}Q^{(1-)}R_{0}^{(1-)}Q^{(1-)}H_{x}Q^{(1-)}R_{0}^{(1-)}Q^{(1-)}H_{r}^{(-)}P+,\\
 & + & PH_{rr}^{(2+)}Q^{(2+)}R^{(2+)}Q^{(2+)}H_{rr}^{(2+)}P+\nonumber \\
 & + & PH_{rr}^{(2-)}Q^{(2-)}R^{(2-)}Q^{(2-)}H_{rr}^{(2-)}P\nonumber 
\end{eqnarray}
which represents the approximate effective Hamiltonian acting in the
subspace with the fixed distrubution of electrons between the \emph{d}-
and \emph{l}-systems. For derivation of its individual terms see Appendix
\ref{sec:Expansion-of-the}.

\subsubsection{Magnetic limit}

In the previous Subsection we arrived to the effective Hamiltonian
describing a PTMC in the configuration subspace with fixed distribution
of electrons between the noncorrelated (diamagnetic) ligands and the
correlated (potentially magnetic) \emph{d}-shells of the TMIs. The
purpose of this and subsequent Subsections is to derive the description
of a PTMC in terms of its \emph{d}-shells only. This is analogous
to the description of a mononuclear TMC by its \emph{d}-shell only,
which is precisely the phenomenological crystal field theory\cite{Bethe,Ballhausen,Bersukerbook}
as sequentially derived in the frames of the EHCF approach.\cite{Tch017}
Like there the variables referring to the \emph{l}-system must be
integrated out by taking an average over the approximate ground state
$\left\vert \Phi_{L}\right\rangle $ of the $l$-system. The only
formal requirement for the wave function $\left\vert \Phi_{L}\right\rangle $
is that it is a single determinant one: 
\begin{equation}
\left\vert \Phi_{L}\right\rangle =\bigwedge\limits _{\lambda,\sigma}a_{\lambda\sigma}^{+}\left\vert 0\right\rangle .\label{eq:SCFfucntion}
\end{equation}
 It can be thus either closed shell RHF, ROHF, or the UHF wave functions.
The subscripts $\lambda,\sigma$ refer respectively for the \emph{l}-MO
and the spin projection.

The uncertainty in defining $\left\vert \Phi_{L}\right\rangle $ represents
a problem. In the mononuclear TMC's the definition of the configuration
subspace in terms of the number of \emph{d}-electrons $n_{d}$ was
sufficient to define the effective electrostatic field induced by
the TMI in the ligands, so that the ligands' ground state $\left\vert \Phi_{L}\right\rangle $
could be calculated in this field with sufficient precision. In a
PTMC the situation at this point is not that clear since electrons
are allowed to travel between the \emph{d}-shells of different TMI's
and such a redistribution produces too strong perturbation of the
\emph{l}-system to be accomodated by a single single-determinant wave
function. For that reason we have to make an additional assumption,
conforming, however, to our purpose of describing magnetic interactions.
We assume that there exists a preferred distribution $\left\{ n_{d}\left(i\right)\right\} $
of electrons among the TMI's in the PTMC. Fixing this distribution
defines the Coulomb field felt by the electrons in the ligands and
thus $\left\vert \Phi_{L}\right\rangle $ can be calculated.

To formalize this move, which in fact reduces the description of the
PTMC to the model configuration subspace where only the local electronic
motions are allowed we define the projection operator acting in the
subspace of the $n_{d}$-electronic states spanned by all possible
determinants formed from AOs of the \emph{d}-system with the fixed
distribution of \emph{d}-electrons: 
\begin{equation}
\mathfrak{P}=\mathbb{O}\otimes I_{l},\label{eq:MMCTProjector}
\end{equation}
 where $I_{l}$ is the identity operator acting in the space of configurations
of the \emph{l}-system and $\mathbb{O}$ projects the configurations
of the \emph{d}-system on the subspace with the fixed numbers of electrons
in each of the \emph{d}-shells as formally defined by eq. (\ref{eq:d-projector})
in Section \ref{sec:Resolvents}. The operator $\mathfrak{Q}$ is
complementary to $\mathfrak{P}:$ $\mathfrak{P}+\mathfrak{Q}=\mathfrak{1}$.

With so defined projection operators we perform a further Löwdin partition
as applied to the effective Hamiltonian $H^{\mathrm{eff}}(E)$ eq.
(\ref{eq:EffectiveHamiltonian}) which results in another effective
Hamiltonian:

\begin{equation}
\mathfrak{H}(E,\mathfrak{E})=\mathfrak{P}H^{\mathrm{eff}}\left(E\right)\mathfrak{P}+\mathfrak{P}H^{\mathrm{eff}}\left(E\right)\mathfrak{Q}\mathfrak{R}(\mathfrak{E})\mathfrak{Q}H^{\mathrm{eff}}\left(E\right)\mathfrak{P}\label{eq:HamiltonianMagneticSubspace}
\end{equation}
 The resolvent is naturally of the form: 
\begin{equation}
\mathfrak{R}(\mathfrak{E})=[\mathfrak{QE}-\mathfrak{Q}H^{\mathrm{eff}}\mathfrak{Q}]^{-1}\label{eq:MMCTResolvent}
\end{equation}
 The action of the projection operators $\mathfrak{Q}$ and $\mathfrak{P}$
on the terms of the effective Hamiltonian $H^{\mathrm{eff}}(E)$ eq.
(\ref{eq:EffectiveHamiltonian}) is: 
\begin{equation}
\begin{array}{c}
\mathfrak{P}PH_{0}P\mathfrak{P}+\mathfrak{P}PH_{x}P\mathfrak{P}\neq0\\
\mathfrak{P}PH_{rr}P\mathfrak{P}\neq0\\
\mathfrak{P}PH_{0}P\mathfrak{Q}=\mathfrak{Q}PH_{0}P\mathfrak{P}=0\\
\mathfrak{P}PH_{rr}P\mathfrak{Q}\neq0;\mathfrak{Q}PH_{rr}P\mathfrak{P}\neq0\\
\mathfrak{Q}H^{\mathrm{eff}}\mathfrak{Q}=\mathfrak{Q}P\left(H_{0}+H_{rr}+H_{x}\right)P\mathfrak{Q}\neq0
\end{array}\label{MagneticProjectorAction}
\end{equation}
 Only the nonvanishing term $\mathfrak{P}PH_{rr}P\mathfrak{Q}$ and
its hermitean conjugate admix the states with charge transfer between
the \emph{d}-shells \emph{i.e.} metal to metal charge transfer states
(hereinafter abbreviated as MMCT) laying in the <<outer>> configuration
space $\mathrm{Im}P\mathfrak{Q}$ to the states in our <<model>>
space $\mathrm{Im}P\mathfrak{P}$.

\subsubsection{Ligand integration and ligand polarization }

Now we are almost ready to perform the final moves to eliminate the
\emph{l}-system variables. In variance with our original way\cite{Tch017}
of reducing the effective Hamiltonian of the type of eqs. (\ref{eq:EffectiveHamiltonian})
and (\ref{eq:HamiltonianMagneticSubspace}) acting in the restricted
configuration space to one describing the mononuclear TMC in terms
of its only \emph{d}-shell which led us to the formal equivalent of
the crystal field theory accompanied by the reliable estimates of
the splitting parameters induced by this field\cite{Tch017} the situation
in the PTMC's is more complicated since it requires more complete
dressing procedure for the \emph{d}-electrons which eventually includes
the effects of polarization of the ligand sphere (diamagnetic media
in which the TMI's are immerced) which accompany the virtual electron
transfers from one \emph{d}-shell to another. Such processes did not
exist in the mononuclear case and for that reason had been originally
neglected. It has been, however, explicitly shown that even in the
mononuclear setting the polarization effects are responsible for the
known nephelauxetic effect,\cite{Joergensen1964,Jorgensen} which
is the renormalization of the intrashell Coulomb (\emph{i.e.} two-electronic)
matrix elements due to indirect interaction between the fluctuations
of the electronic density in the \emph{d}-shell through the polarizations
they induce in the ligands.\cite{Tch083} One may be sure that the
partition described by eq. (\ref{eq:EffectiveHamiltonian}) \emph{does
not} lead to the description of the TMC by the wave function of the
form eq. (\ref{eq:WaveFunction}) rather by a description through
a superposition of the functions of that form each corresponding to
fixed number of electrons in the subsystems, but still entangled.
In fact the operator $P$ projects out the states with the charge
transfer between the $d$- and $l$-system, but does not assure the
functional product form of eq. (\ref{eq:WaveFunction}). Products
of different states of the $d$- and $l$-systems satisfying only
the condition of the fixed number of the electrons in each of the
singled out groups enter in the expansion of the true ground state
of the effective Hamiltonian eq. (\ref{eq:EffectiveHamiltonian}).
The form eq. (\ref{eq:WaveFunction}) is a kind of self consistent
field approximation to it.\cite{McWeenybook} In order to improve
this description one has to perform one more projection namely to
the subspace of the products where all possible states of the $l$-system
are replaced by its ground state. This is done as previously with
use of the Löwdin projection using further projection operators. Let
$\mathcal{P}$ be the projection operator on the ground state $\left\vert \Phi_{L}\right\rangle $
of the $l$-system: 
\begin{equation}
\mathcal{P}=I_{d}\otimes\left\vert \Phi_{L}\right\rangle \left\langle \Phi_{L}\right\vert \label{eq:l-GroundStateProjector}
\end{equation}
 complemented by the Kronecker product with $I_{d}$ - the identity
operator with respect to the variables of the \emph{d}-system. The
complementary projection operator is $\mathcal{Q}=P\mathfrak{P}-\mathcal{P}$.
Then applying the Löwdin partition as in the previous Subsections
we arrive to the effective operator $\mathcal{H}^{eff}(E,\mathfrak{E},\Omega)$
which acts in the subspace of the configurations where the $l$-system
is fixed in its ground state $\left\vert \Phi_{L}\right\rangle $:
\begin{equation}
\mathcal{H}^{\mathrm{eff}}(E,\mathfrak{E},\Omega)=\mathcal{P}\mathfrak{H(E)}\mathcal{P}+\mathcal{P}\mathfrak{H(E)}\mathcal{QR}(\Omega)\mathcal{Q}\mathfrak{H(E)}\mathcal{P}.\label{eq:PolarizationResolvent}
\end{equation}
 The Hamiltonian eq. (\ref{eq:PolarizationResolvent}) does not affect
the variables in the \emph{l}-system as described by its ground state
$\left\vert \Phi_{L}\right\rangle $. Thus the effective Hamiltonian
for the \emph{d}-system can be easily obtained by dropping the bra-
$\left\langle \Phi_{L}\right\vert $ on the right and the ket $\left\vert \Phi_{L}\right\rangle $
on the left in eq. (\ref{eq:PolarizationResolvent}) (namely in this
order). Then the remaining averages over the wave function $\left\vert \Phi_{L}\right\rangle $
yield the sought effective Hamiltonian for the \emph{d}-system only
(the variables related to the \emph{l}-system are integrated out).
In the subsequent Sections we identfy in turn the individual contributions
to it although the precise derivation of corresponding expressions
will be kept in the respective Appendices. Further moves are stipulated
by the projection to the single determinant ground state of the $l$-system
as given by eq. (\ref{eq:PolarizationResolvent}). With use of the
above generalized operators one can easily write the excited states
relative to the single determinant ground state.

For completeness we calculate the resolvent part of the operator eq.
(\ref{eq:PolarizationResolvent}). We notice that for the ground state
of the vanishing total spin (closed shell) assumed hereinafter the
following holds:

\begin{eqnarray*}
\mathcal{P}H_{x}\mathcal{P} & = & 0;\\
\mathcal{P}H_{x}\mathcal{Q} & \neq & 0
\end{eqnarray*}
 and calculate the second expression (or hermitean conjugate to it).
It is also true that $\mathcal{P}H_{rr}^{(0)}\mathcal{Q}\neq0$ and
thus the operator $H_{rr}^{(0)}$ contributes to the second order
according to 
\[
\begin{array}{c}
\mathcal{P}H_{rr}^{(0)}\mathcal{R}(\Omega)H_{rr}^{(0)}\mathcal{P}=\\
\mathcal{P}H_{rr}^{(0)}(\mathcal{R}_{S}(\Omega)+\mathcal{R}_{T}(\Omega))H_{rr}^{(0)}\mathcal{P}=\\
\mathcal{P}H_{rr}^{(0)}\mathcal{R}_{S}(\Omega)H_{rr}^{(0)}\mathcal{P}+\mathcal{P}H_{rr}^{(0)}\mathcal{R}_{T}(\Omega)H_{rr}^{(0)}\mathcal{P}
\end{array}
\]
$\mathcal{R}_{S}(\Omega)$ and $\mathcal{R}_{T}(\Omega)$ contain
respectively the projection operators to the singlet and triplet subspaces
in the $\mathrm{Im}\mathcal{Q}$ subspace. The average $\left\langle \Phi_{L}\left|H_{rr}^{(0)}\mathcal{R}(\Omega)H_{rr}^{(0)}\right|\Phi_{L}\right\rangle $
contributes to the effective Hamiltonian for the $d$-electrons only.

\section{Specific contributions to the \emph{d}-system effective Hamiltonian.\label{sec:Specific-contributions-to}}

As shown in the previous Section the contributions to the effective
Hamiltonian for the \emph{d}-system of a PTMC stem from averaging
eq. (\ref{eq:PolarizationResolvent}) over $\left\vert \Phi_{L}\right\rangle $.
As we mentioned in the previous Subsection the leading terms are those
which appear from the first term of eq. (\ref{eq:PolarizationResolvent})
or equivalently the average of the entire Hamiltonian eq. (\ref{eq:HamiltonianMagneticSubspace}).
In the subsequent Subsections we consider them one by one.

\subsection{Effective crystal field}

The simplest contribution comes from the averaging the first term
in eq. (\ref{eq:HamiltonianMagneticSubspace}) directly contributing
to the first term of eq. (\ref{eq:PolarizationResolvent}) over $\left\vert \Phi_{L}\right\rangle $.\cite{Tch017}
The latter in its turn contains two contribution: the average of the
Coulomb interaction over the ground state of the \emph{l}-system $\left\vert \Phi_{L}\right\rangle $
which trivially (see Appendix \ref{sub:Ionic-contribution-to}) yields
the sum of the ionic/intraatomic contributions to the effective crystal
field felt by the \emph{d}-shells of the individual TMIs in a PTMC.
It is consistently shown that the ionic contribution does not exceed
10\% of the observed splitting.\cite{Tch075,Tch029,Tch030,Tch033,Tch037}

Nontrivial (and eventually the dominant) part of the latter is given
by the average of the term $H_{rr}^{(0)}$ of the second order in
$H_{r}$ originally performed while deriving the EHCF method for mononuclear
TMCs.\cite{Tch017} Those derivation is generalized by treating individually
the terms diagonal with respect to the TMI's indices $ij$ as shown
in Appendix \ref{sub:Covalent-contribution-to}: 
\begin{equation}
\begin{array}{c}
\left\langle \Phi_{L}\right\vert \mathfrak{P}PH_{rr}^{(0)}P\mathfrak{P}\left\vert \Phi_{L}\right\rangle =\sum_{j}V_{\mathrm{CF}}^{\mathrm{cov}}(j)=\sum_{j}\sum\limits _{\mu\nu}\sum\limits _{\kappa\tau}\beta_{\mu\kappa}(j)\beta_{\nu\kappa}(j)\times\\
\left\{ n_{\kappa}D^{(1+)}(j\kappa)d_{\nu\tau}^{+}(j)d_{\mu\tau}(j)-(1-n_{\kappa})D^{(1-)}(j\kappa)d_{\mu\tau}^{+}(j)d_{\nu\tau}(j)\right\} 
\end{array}\label{eq:GeneralizedCrystalField}
\end{equation}
 This operator due to projection operator $\mathfrak{P}$ represents
precisely as denoted the sum of the covalent contributions to the
effecitve crystal fields felt by the \emph{d}-shells of the individual
TMIs in the PTMC. Then the overall result of the averaging of $\mathfrak{P}H^{\mathrm{eff}}\mathfrak{P}$
reads: 
\begin{eqnarray}
 & \left\langle \Phi_{L}\left\vert \mathfrak{P}P\left(H_{0}+H_{rr}+H_{x}\right)P\mathfrak{P}\right\vert \Phi_{L}\right\rangle \nonumber \\
= & \sum_{i}\left\{ H_{d}(i)+V_{\mathrm{CF}}^{\mathrm{ion}}(i)+V_{\mathrm{CF}}^{\mathrm{at}}(i)+V_{\mathrm{CF}}^{\mathrm{cov}}(i)\right.\label{eq:PolynuclearEHCF}\\
 & \left.-2\sum_{\mu l}K_{\mu l}(i)\hat{S}_{\mu}(i)\left\langle \Phi_{L}\left\vert \hat{S}_{l}(i)\right\vert \Phi_{L}\right\rangle \right\} \nonumber 
\end{eqnarray}
 The summation in the last term extends to the \emph{d}-AOs $\mu$
and \emph{l}-AOs \emph{l} of the \emph{i}-th TMI. It appears only
if the spin density on the $l$-th AO of the $i$-th atom is nonvanishing.
It is possible for the ROHF and UHF functions $\left|\Phi_{L}\right\rangle $
only and it is new as compared to the original EHCF\cite{Tch017}
where only the RHF form of $\left|\Phi_{L}\right\rangle $ was allowed.
However, hereinafter we concentrate on the systems in which in the
zero approximation the ligands are diamagnetic, \emph{i.e.} have no
magnetic moments/spin densities in the \emph{l}-system (and are thus
representable by the RHF function) so that this term further is not
going to reappear.

\subsection{Magnetic contributions}

\begin{figure}
\includegraphics{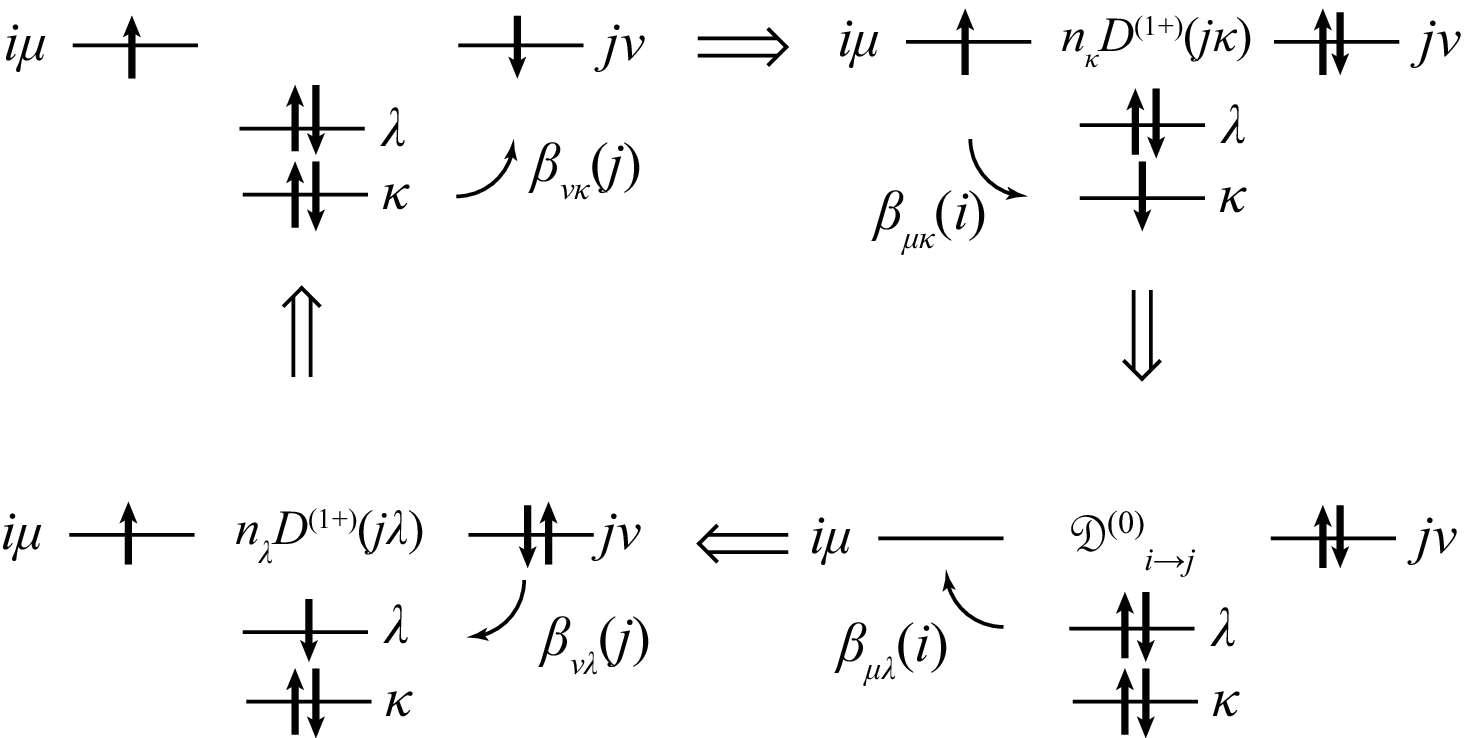}

\caption{Sequence of configurations relevant to the matrix elements of type
(a) as contributing to the effective exchange interaction in the case
(i). The configuration in the upper left corner belongs to the model
subspace $\mathrm{Im}P\mathfrak{P}$. The configurations in the upper
right and lower left corners belong to the outer subspace $\mathrm{Im}Q^{(1+)}$
- that formed from the model space configurations by the LMCT states/processes,
that in the lower right corner belongs to the outer subspace $\mathrm{Im}\mathfrak{Q}$
- that formed from the model space configurations by the MMCT states/processes.
Transitions between the subspaces are indicated by the arrows, whereas
the specific electron transfers by the arc arrows. The contributions
to the exchange parameter are obtained as products of the quantities
indicated in the Figure. The case (i) is characterized by the choice
of the values of the energy denominators: namely those with the lowering
spin in the both \emph{i}-th and \emph{j}-th \emph{d}-shells.\label{fig:TypeAi} }
\end{figure}

\begin{figure}
\includegraphics{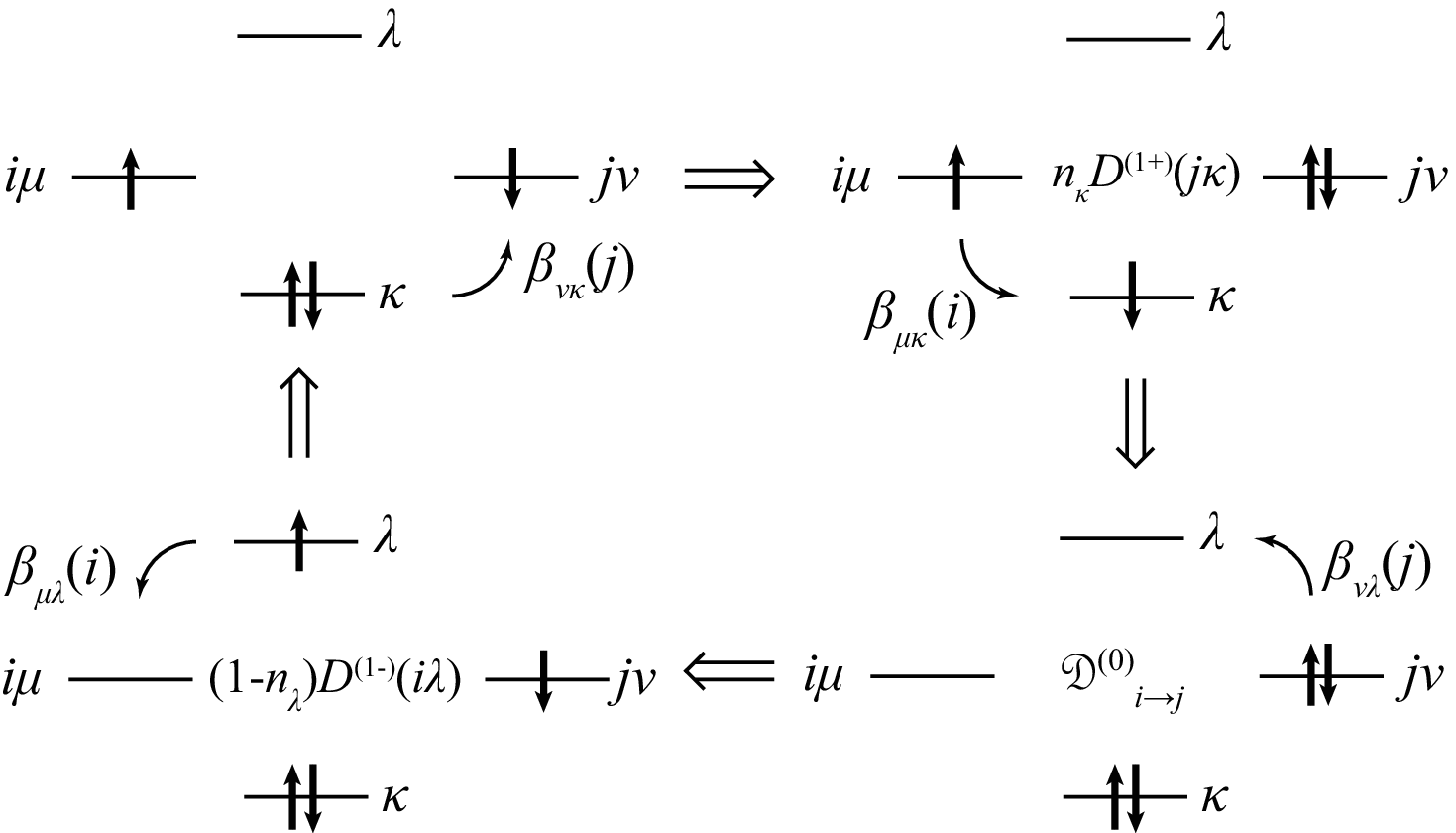}\caption{Sequence of configurations relevant to the matrix elements of type
(b) as contributing to the effective exchange interaction in the case
(i). The configurations in the upper right and lower left corners
belong to the outer subspaces $\mathrm{Im}Q^{(1\pm)}$ - that formed
from the model space configurations by the LMCT states/processes.
Other notation is as in Fig. \ref{fig:TypeAi}. \label{fig:TypeBi} }
\end{figure}

\begin{figure}
\includegraphics{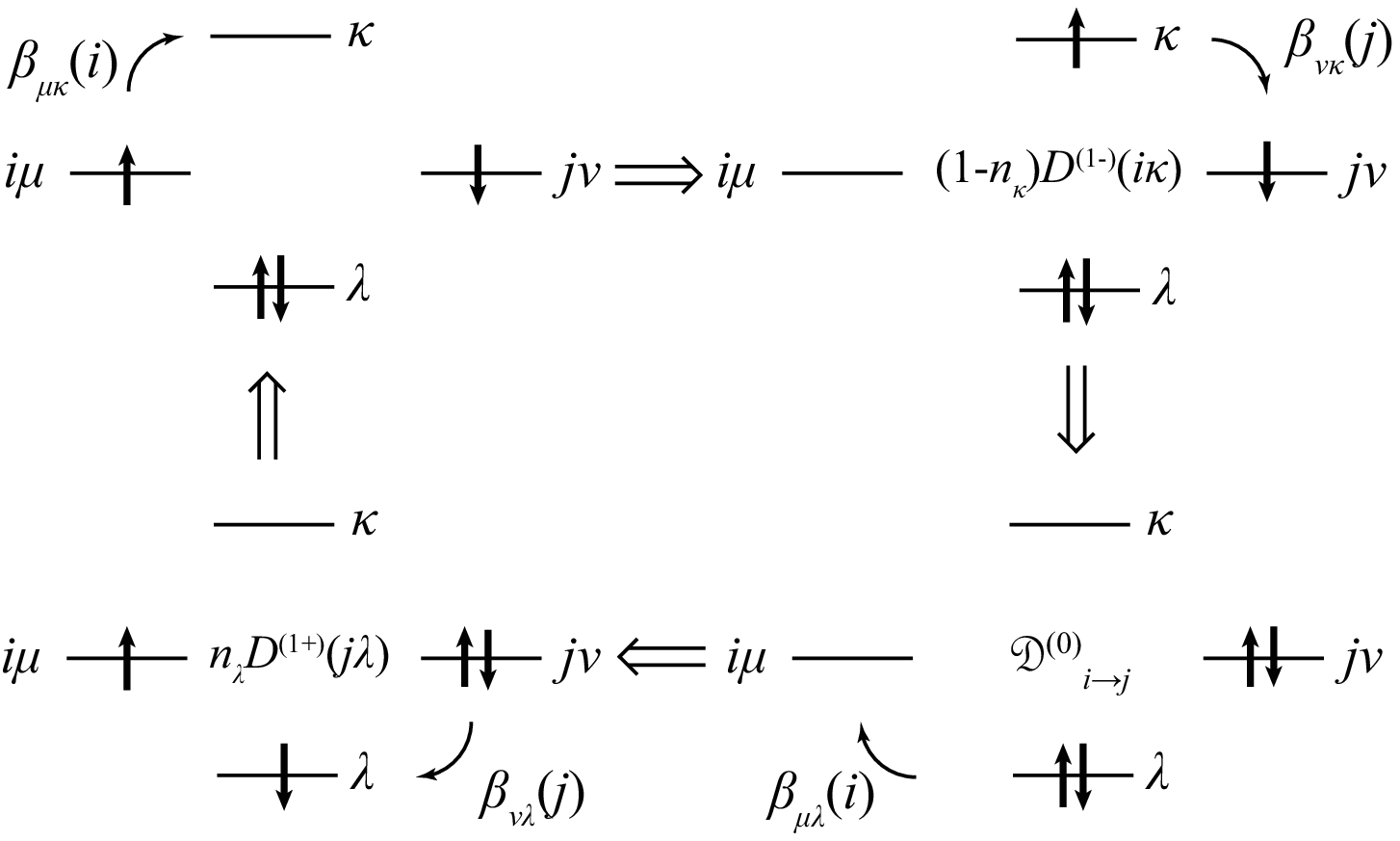}\caption{Sequence of configurations relevant to the matrix elements of type
(c) as contributing to the effective exchange interaction in the case
(i). The configurations in the upper right and lower left corners
belong to the outer subspaces $\mathrm{Im}Q^{(1\mp)}$ - that formed
from the model space configurations by the LMCT states/processes.
Other notation is as in Fig. \ref{fig:TypeAi}. \label{fig:TypeCi} }
\end{figure}

\begin{figure}
\includegraphics{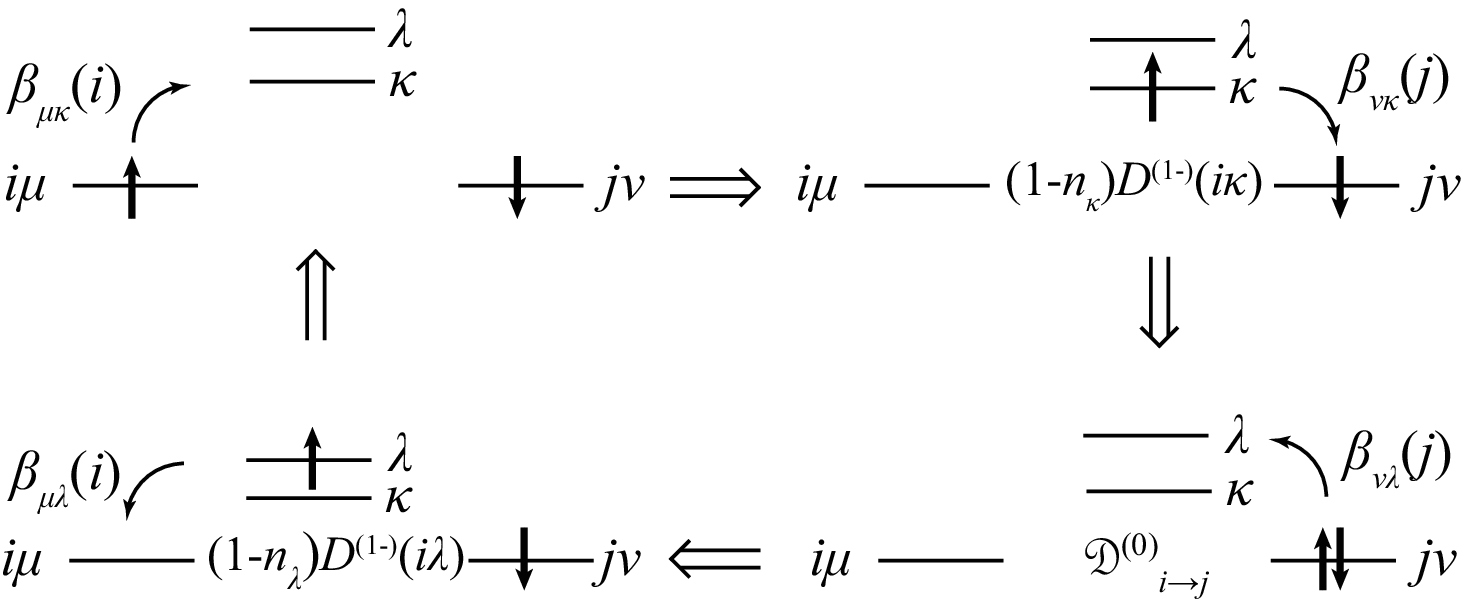}\caption{Sequence of configurations relevant to the matrix elements of type
(d) as contributing to the effective exchange interaction in the case
(i). The configurations in the upper right and lower left corners
belong to the outer subspaces $\mathrm{Im}Q^{(1-)}$ - that formed
from the model space configurations by the LMCT states/processes.
Other notation is as in Fig. \ref{fig:TypeAi}. \label{fig:TypeDi} }
\end{figure}

\begin{figure}
\includegraphics{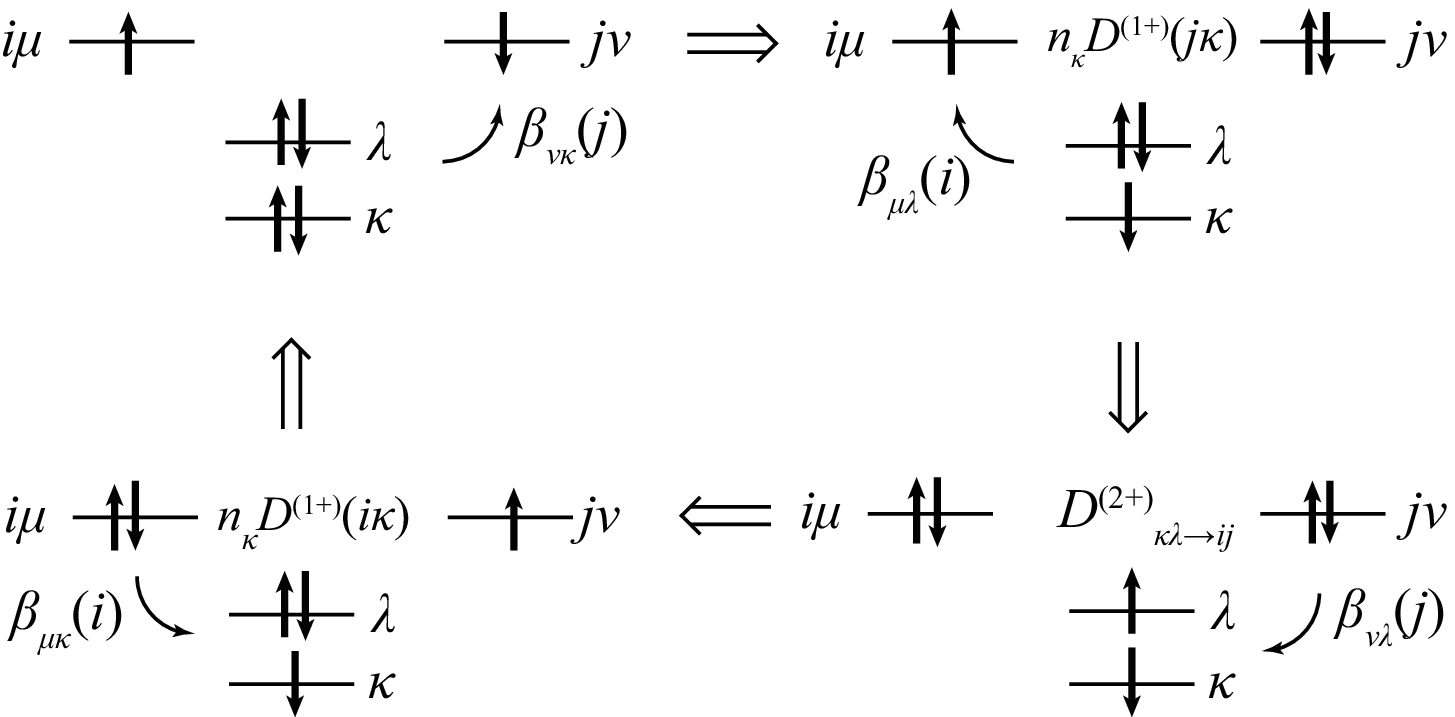}\caption{Sequence of configurations relevant to the matrix elements of type
(e) as contributing to the effective exchange interaction in the case
(i). The configurations in the upper right and lower left corners
belong to the outer subspace $\mathrm{Im}Q^{(1+)}$ - that formed
from the model space configurations by the LMCT states/processes,
that in the lower right corner belongs to the outer subspace $\mathrm{Im}Q^{(2+)}$
- that formed from the model space configurations by the (LM)$^{2}$CT
states/processes. Other notation is as in Fig. \ref{fig:TypeAi}.
\label{fig:TypeEi} }
\end{figure}

In the previous Subsection we rederived the second order contribution
of the one-electron hopping between the \emph{d}- and \emph{l}-systems
to the effective Hamiltonian. Its part which is diagonal with respect
to the TMI's indices $i$ and \emph{j} (or, equivalently, one acting
within the configuration subspace $\mathrm{Im}\mathfrak{P}$) resulted
as expected in the sum of the crystal field contributions for individual
TMIs. The $ij$-off-diagonal part of $H_{rr}^{(0)}$ (that according
to the formal expression $\mathfrak{P}H_{rr}^{(0)}\mathfrak{Q}$ coupling
the configurations in the subspaces $\mathrm{Im}\mathfrak{P}$ and
$\mathrm{Im}\mathfrak{Q}$) contributes to the effective operator
for the \emph{d}-system according to: 
\begin{equation}
\begin{array}{c}
\left\langle \Phi_{L}\left|\mathfrak{P}H_{rr}^{(0)}\mathfrak{Q}\mathfrak{R}(\mathfrak{E})\mathfrak{Q}H_{rr}^{(0)}\mathfrak{P}\right|\Phi_{L}\right\rangle =\\
\left\langle \Phi_{L}\left|\mathfrak{P}H_{rr}^{(0)}\mathfrak{Q}\left[\mathfrak{R}_{S}(\mathfrak{E})+\mathfrak{R}_{T}(\mathfrak{E})\right]\mathfrak{Q}H_{rr}^{(0)}\mathfrak{P}\right|\Phi_{L}\right\rangle 
\end{array}\label{eq:MMCTPartition}
\end{equation}
 by which the MMCT configurations are projected out. Formally it is
nothing but the average of the second term in eq. (\ref{eq:HamiltonianMagneticSubspace}).
Eventually it represents the leading magnetic contribution. Details
of the derivation are given in Appendix \ref{sub:Effective-one-electron-hopping}.
The corresponding configurations are shown in Figs. \ref{fig:TypeAi}
- \ref{fig:TypeDi}. They describe the situation which is conveniently
described as effective one-electron transfers between different \emph{d}-shells.
However, it has been shown\cite{WeiheGuedelToftlund} that other processes
having nothing to do with transfers of electrons \emph{between} interacting
\emph{d}-shells rather those which can be described as correlated
electron transfers between the \emph{l}-system and the \emph{d}-shells
in which two electrons of the opposite spin projection are excited
from a single \emph{l}-MO to two \emph{d}-shells under consideration.
The correlations in this case is as well of <<kinematic>> nature
and thus they contribute predominantly antiferromagnetically. The
corresponding configurations are shown in Fig. \ref{fig:TypeEi}.
these contributions appear as a result of averaging the terms of the
second order with respect to the ionizing components $H_{rr}^{(2\pm)}$
in eq. (\ref{eq:EffectiveHoppingComponents}) of Appendix \ref{sec:Expansion-of-the}.

The terms eqs. (\ref{eq:CorrelatedIonizationSpecific}), (\ref{eq:aaa-1})
must be summed with the spin dependent factors assembled in Table
\ref{tab:Spin-dependent-factors}of Appendix \ref{sub:Spin-operator-contributions}
and accroding to one of the four {}``cases'' (i) - (iv) identified
by Weihe and Güdel\cite{WeiheGuedel1997} to take into account the
effect of the different variations of the spins in the \emph{d}-shells
involved in the electron transfer processes up on the contributions
to the effective exchange between these \emph{d}-shells. As one can
see from Table \ref{tab:Spin-dependent-factors} the terms with different
variation of the local spins contribute with different sign. The classification
to the {}``cases'' (i) - (iv) is based on the assumption that the
transfer of an electron involving a half-filled orbital always result
in a reduction of the local spins by $1/2$ in either of the involved
\emph{d}-shells irrespective to the sense of the transfer. If either
of the involved orbitals is either completely occupied or empty the
electron transfer process may result either in increase or decrease
of the local spin by $1/2$ in the \emph{d}-shell where such an orbital
occurs. This results in only one contribution for the case (i), two
contributions for the cases (ii) and (iii), and four contributions
for the case (iv). These contributions enter with numerical unpairity
factors specific for each case and with the signs specific for each
combination of the possible variations of the local spins. The values
of the energy denominators ($D^{(1\pm)},D^{(2\pm)},\mathfrak{D}^{(0)}$)
referring to the intermediate states in the outer configuration subspaces
must be taken according to the variation of the local spins specific
for each contribution (that is to be higher by the intrashell exchange
energy when the spin goes down and to be smaller by the same quantity
when the spin goes up). That structure of the contributions and of
their combinations entering the final expressions for the exchange
parameters moved the authors\cite{WeiheGuedelToftlund} to expand
the denominators' products against the intraatomic exchange energies.
Due to alternating signs of the contributions corresponding to the
different conbinations of the intermediate local spins the terms of
the lowest nonvanishing order survive for each of the cases (i) -
(iv) which are respectively of the zeroth power for the case (i),
of power one for the cases (ii) and (iii) and of the second power
for the case (iv). This is how the expressions for the cases (i) -
(iv) have been derived previously.\cite{WeiheGuedelToftlund} We do
not use these expansions and employ {}``exact'' values of the energy
denominators. 

Our derivation allows an additional classification according to the
types of the terms in the sum over $\kappa\lambda$ which run through
the \emph{l}-MOs. These types (a) - (d) formally defined by eq. (\ref{eq:aaa})
correspond to the four possible transfer paths shown in Figs. \ref{fig:TypeAi}
- \ref{fig:TypeDi} and respectively involving two occupied, two empty,
and one empty and one occupied \emph{l}-MOs in various orders. They
describe contributions of the states in the MMCT configutation subspace
and thus contain the energy denominator $\mathfrak{D}_{i\rightarrow j}^{(0)}$. 

The values of these factors are generalized and summarized in Table
\ref{tab:Multipliers }. Each of these matrix elements contains the
following expression:

\begin{equation}
\beta_{\nu\kappa}(j)\beta_{\mu\kappa}(i)\beta_{\mu\lambda}(i)\beta_{\nu\lambda}(j)\left[\mathfrak{D}_{i\rightarrow j}^{(0)}+n_{\kappa}n_{\lambda}D_{\kappa\lambda\rightarrow ij}^{(2+)}\right]\label{eq:**}
\end{equation}
as a common factor. The first term in square brackets corresponds
to an electron transfer from the \emph{i}-th \emph{d}-shell to the
\emph{j}-th one. The second term comes from the process of spin-correlated
transfer of two electrons with opposite spin projections from the
occupied \emph{l}-MOs to the two \emph{d}-shells by this effectuating
coupling between them.

\begin{table}
\caption{Generalized factors of the matrix elements contributing to the exchange
parameters. \label{tab:Multipliers }}

\begin{tabular}{|c|c|}
\hline 
\emph{a } & $n_{\kappa}n_{\lambda}D^{(1+)}(j\lambda)D^{(1+)}(j\kappa)$\tabularnewline
\hline 
\emph{b } & $-n_{\kappa}\left(1-n_{\lambda}\right)D^{(1+)}(j\kappa)D^{(1-)}(i\lambda)$\tabularnewline
\hline 
\emph{c } & $-\left(1-n_{\kappa}\right)n_{\lambda}D^{(1-)}(i\kappa)D^{(1+)}(j\lambda)$\tabularnewline
\hline 
\emph{d } & $\left(1-n_{\lambda}\right)\left(1-n_{\kappa}\right)D^{(1-)}(i\kappa)D^{(1-)}(i\lambda)$\tabularnewline
\hline 
\emph{e} & $n_{\kappa}n_{\lambda}D^{(1+)}(j\lambda)\left[D^{(1+)}(i\kappa)+D^{(1+)}(j\lambda)\right]$\tabularnewline
\hline 
\emph{f} & $\left(1-n_{\lambda}\right)\left(1-n_{\kappa}\right)D^{(1-)}(i\lambda)\left[D^{(1-)}(i\kappa)+D^{(1-)}(j\lambda)\right]$\tabularnewline
\hline 
\end{tabular} 
\end{table}

In the previous Sections we were able to basically rederive and generalize
the perturbative formulae for the contributions to the exchange in
a form suitable for programming. They are also complemented by additional
terms stemming from various processes involving polarization of the
\emph{l}-system. As expected the terms are numerous. However, additional
considerations allow to range these terms in order of their importance.
At the present first step we concentrate on the estimates of the contributions
stemming from the electron transfers which are formally of the fourth
power with respect to the one-electron hopping integrals $\beta$'s.

\section{Details of implementation \& calculation results\label{sec:Calculations-results}}

The formulae for the effective exchange parameters derived in the
previous Sections do not require any additional quantities except
those which are already calculated in the context of the EHCF method.\cite{Tch017}
These are the \emph{l}-MO LCAO expansion coefficients, respective
orbital energies, \emph{d-}AO-\emph{l}-MO resonance integrals $\beta_{\mu\lambda}\left(i\right)$,
\emph{etc.} We implemented the derived formulae as a program suite
accepting standard quantum chemical input (molecular composition and
geometry) using the GEOMO package\cite{GEOMO} (QCPE No 290) as a
source of the subroutines for performing the calulations of the molecular
integrals and performing semi-empirical SCF MO LCAO procedures for
the \emph{l}-system. The package has been tested agaist the compounds
of the Cr(III)OCr(III) family\cite{OCrNH3,GAMTUJ,VIDTIL,ZUVMIM} known
as <<basic rhodo>> compounds which have been synthetized by S.M.
Jorgensen 130 years ago.\cite{Jorgensen1882} The parameterization
procedure follows general EHCF methodology and will be described elsewhere.

We performed a series of calculations for those compounds studied
in the phenomenological setting where the structural data were available.
The results relative to the $\mu$-oxo bridged Cr(III) dimers are
given in Table \ref{tab:Cr-III-dimers}. 
\begin{table}
\caption{Calculated and experimental values of the exchange parameters in $\mu$-oxo
bridged Cr(III) dimers.\label{tab:Cr-III-dimers} }

\begin{tabular}{|c|c|c|c|}
\hline 
Compound (CCSD Code)  & $J_{\mathrm{exp}}$, cm$^{-1}$ Ref. \cite{MossinWeihe}  & $J_{\mathrm{calc}}$, cm$^{-1}$  & Geometry source \tabularnewline
\hline 
\hline 
{[}(NH$_{3}$)$_{5}$CrOCr(NH$_{3}$)$_{5}${]}$^{4+}$  & 450  & 408  & \cite{OCrNH3} \tabularnewline
\hline 
GAMTUJ  & 510  & 424  & \cite{GAMTUJ} \tabularnewline
\hline 
VIDTIL  & 100  & 190  & \cite{VIDTIL} \tabularnewline
\hline 
ZUVMIM  & 60  & 150  & \cite{ZUVMIM} \tabularnewline
\hline 
\end{tabular}
\end{table}

As one can see the order of magnitude of the exchange parameters in
this series of compounds is correctly reproduced as are their trends
dependent on the chemical composition and bridge geometry (basic erhythro-compound
{[}(NH$_{3}$)$_{5}$CrOCr(NH$_{3}$)$_{5}${]}$^{4+}$ and GAMTUJ
have linear bridge geometry whereas VIDTIL and ZUVMIM are bent with
the $\widehat{\mathrm{CrOCr}}$ angle in the range 128$^{\circ}$$\div$132$^{\circ}$)
, although the amplitude of the angular dependence as obtained in
the calculation is somewhat smaller than the experiemental one. Analogous
calculations performed for the $\mu$-oxo bridged Fe(III) dimers (Table
\ref{tab:Fe-III-dimers}) 
\begin{table}
\caption{Calculated and experimental values of the exchange parameters in $\mu$-oxo
bridged Fe(III) dimers.\label{tab:Fe-III-dimers} }

\begin{tabular}{|c|c|c|c|}
\hline 
Compound (CCSD Code)  & $J_{\mathrm{exp}}$, cm$^{-1}$ Ref. \cite{WeiheGuedel1998}  & $J_{\mathrm{calc}}$, cm$^{-1}$  & Geometry source \tabularnewline
\hline 
\hline 
DIBXAN  & 238  & 240  & \cite{DIBXAN} \tabularnewline
\hline 
VABMUG  & 264  & 256  & \cite{VABMUG} \tabularnewline
\hline 
PYCXFE  & 214  & 118  & \cite{PYCXFE} \tabularnewline
\hline 
COCJIN deprotonated  & 242  & 110  & \cite{COCJIN} \tabularnewline
\hline 
COCJIN  &  & 30  & \cite{COCJIN10} \tabularnewline
\hline 
\end{tabular}
\end{table}
show similarly reasonable agreement between the experimental and calculated
values of the exchange constants.

The most remarkable feature we addressed in this context is the effect
of protonation of the oxo-bridge upon the magnitude of the exchange
constant. Comparing the calculated values in the last two lines of
Table \ref{tab:Fe-III-dimers} we see that the protonation as expected
breaks at least one of the superexchange paths going through the oxo-bridge
which as expected as well significantly reduces the magnitude of the
effective exchange parameter.

\section{Conclusion}

By this we extend our work \cite{Tch017} of the year 1992 devoted
to calculating the intrashell excitations in the \emph{d}-shells of
coordination compounds of the first transition metal row, which resulted
in the Effective Hamiltonian Crystal Field (EHCF) method, to their
polinuclear analogs in order to assure the description of several
open \emph{d}-shells and of magnetic interactions of the effective
spins residing in these shells. This is a challenging task since it
requires improving the precision of \emph{ca.} 1000 $\mathrm{cm}{}^{-1}$
(that of describing the excitation energies of the single \emph{d}-shells
by the already well successful EHCF method) to the that of \emph{ca.}
10 $\div$ 100 $\mathrm{cm}{}^{-1}$ characteristic for the energies
required to reorient the spins \emph{i.e.} eventually by two orders
of magnitude. This is performed within the same paradigm as used for
the EHCF method: the concerted usage of the McWeeny's group-function
approximation and the Löwdin partition technique. These are used to
develop the effective description of the \emph{d}-system of the polynuclear
complexes composed of several \emph{d}-shells including the working
formulae for the exchange parameters between the \emph{d}-shells belonging
to different transition metal ions. These formulae are implemented
in the package MagAîxTic and tested against a series of binuclear
complexes of trivalent cations featuring $\mu$-oxygen superexchange
paths in order to confirm the reproducibility of the trends in the
series of values of exchange parameters as well as the magnitude of
these parameters for the compounds differing by the chemical substitutions
and other details of composition and structure. The results of calculations
are in a reasonable agreement with available experimental data and
other theoretical methods.

\section*{Acknowledgments}

This work is supported by RFBR through the grants Nos 10-03-00155,
13-03-00406, and 13-03-90430. 

\appendix

\section{Hamiltonian contributions\label{sec:Hamiltonian-contributions}}

\subsection{Subsystem's Hamiltonians\label{sub:Subsystem's-Hamiltonians}}

\subsubsection{Hamiltonian for the \emph{d}-System \label{sub:d-System-Hamiltonian}}

The bare Hamiltonian for the \emph{d}-system reads:

\[
\sum_{i}H_{d}(i)+\frac{1}{2}\sum_{i\neq j}V_{\mathrm{Coul}}(ij)
\]
 The Hamiltonians for the individual \emph{d}-shells are taken in
the atomic symmetric approximation dating back to Refs. \cite{Fulde,KleinmanMednick,Oles,OlesStollhoff,Hammermesh}
so that the Coulomb interaction and the exchange splitting of \emph{d}-electrons
are decribed by two parameters ($U_{dd}(i)$ is the one-electron core-attraction
parameter): 
\begin{equation}
H_{d}(i)=U_{dd}(i)\hat{n}_{d}(i)+\frac{1}{2}g_{dd}(i)\hat{n}_{d}(i)\left(\hat{n}_{d}(i)-1\right)-K_{dd}(i)\hat{S}_{d}^{2}(i)\label{eq:d-shellBareHamiltonian}
\end{equation}
where $\hat{n}_{d}(i)$ is the operator of the number of particles,
$\hat{S}_{d}(i)$ is the operator of the total spin, $U_{dd}(i)$
is the one-electron core attraction parameter, $g_{dd}(i)$ is the
average parameter of the Coulomb interaction, $K_{dd}(i)$ is the
average parameter of electronic exchange, all referring to the\emph{
i}-th \emph{d}-shell. The main advantage of this form of the Hamiltonian
is that it preserves a higher $SO(5)$ rather than the actual $SO(3)$
symmetry of the atom so that it remains invariant under arbitrary
5$\times$5 orthogonal transformation of the one-electron \emph{d}-states.
This will be necessary when going to the basis of the eigenstates
of the local crystal field operator.

With such a bare Hamiltonian the state of the \emph{i}-th \emph{d}-shell
is uniquely characterized by two quantum numbers: $n_{d}(i)$ - number
of electrons, we use to build the configuration subspaces and to decompose
the system in parts, and $S_{d}(i)$ - the total spin of the given
$d$-shell. Then the energy of the entire manifold of the states with
given $n_{d}(i)$ and $S_{d}(i)$ is given by: 
\begin{equation}
U_{dd}(i)n_{d}(i)+\frac{1}{2}g_{dd}(i)n_{d}(i)\left(n_{d}(i)-1\right)-K_{dd}(i)S_{d}(i)\left(S_{d}(i)+1\right)\label{eq:d-shellBareEnergy}
\end{equation}
 This form assures for the Hund's rule for the states of the \emph{d}-shells
due to positiveness of the intrashell exchange parameters $K_{dd}(i)$
and for the compliance of the bare spectrum with the Landé interval's
rule as suggested in Refs. \cite{Jorgensen,Joergensen1964}. The values
of the core attraction parameters are taken as implemented in the
EHCF package \cite{Tch017}. Those of the average interaction parameters
are given by Jørgensen as well: 
\begin{eqnarray*}
g_{dd}(i) & = & F_{dd}^{0}(i)-\frac{2}{63}\left(F_{dd}^{2}(i)+F_{dd}^{4}(i)\right)\\
K_{dd}(i) & =\frac{35}{12} & B(i)+\frac{7}{6}C(i)=\frac{5}{84}\left(F_{dd}^{2}(i)+F_{dd}^{4}(i)\right)
\end{eqnarray*}
 - not a very great improvement over Refs. \cite{WeiheGuedel1996,WeiheGuedel1997,WeihGuedel2000,WeiheGuedelToftlund}
where $B(i)$ and $C(i)$ are the Racah parameters for the \emph{i}-th
TMI and $F_{ll^{\prime}}^{k}\left(i\right)$ are the corresponding
Slater-Condon parameters.

\subsubsection{Hamiltonian for the \emph{l}-system}

INDO parameterization for the first row elements has been introduced
in Ref. \cite{PopleBeveridge}. Its extensions to the transition metal
atoms had been proposed in Refs. \cite{ClackHushYandle,BaconZerner,BoehmGleiter}.
Within that setting the main problem was to implement intraatomic
Coulomb and exchange two-electron integrals (Slater-Condon parameters
$F^{k},G^{k}$) since wihtin the INDO (SCF) setting they cover all
necessary intraatomic parameters allowed by symmetry. Rinaldi\cite{GEOMO}
have shown that the intraatomic configuration interaction involving
\emph{s}-, $p$-, and $d$-(sub)shells requires additional intraatomic
integrals. These latter, however, again disappear if the \emph{d}-shells
enter the wave function as direct multipliers eq. (\ref{eq:WaveFunction}).
This is thus the case for the present method. 

Then in the second quantization form the Hamiltonian for the $l$-system
reads: 
\begin{equation}
\begin{array}{c}
H_{l}=\sum\limits _{m,\sigma}(U_{mm}-\sum\limits _{L}V_{ML})m_{\sigma}^{+}m_{\sigma}+\\
+\sum\limits _{L}\sum\limits _{l\in L,\sigma}(U_{ll}-\sum\limits _{L^{\prime}\neq L}V_{LL^{\prime}}-V_{LM})l_{\sigma}^{+}l_{\sigma}+\\
+\sum\limits _{ml,\sigma}\beta_{ml}(m_{\sigma}^{+}l_{\sigma}+h.c.)+\sum\limits _{ll^{\prime},\sigma}\beta_{ll^{\prime}}l_{\sigma}^{+}l_{\sigma}^{\prime}+\\
+\frac{1}{2}\sum_{ll^{\prime},\sigma}(ll^{\prime}\mid l^{\prime\prime}l^{\prime\prime\prime})l_{\sigma}^{+}l_{\tau}^{\prime\prime+}l_{\tau}^{\prime\prime\prime}l_{\sigma}^{\prime};
\end{array}\label{LSystemEffectiveHamiltonianExplicit}
\end{equation}
here\ $l_{\sigma}^{+}$ ($l_{\sigma}$) are the creation (annihilation)
of an electron with the spin projection $\sigma$ on an $l$-AO. First
term in the expression eq. (\ref{LSystemEffectiveHamiltonianExplicit})
describes the interaction of the $4s$- and $4p$-electrons of the
metal ($m=4s,4p_{x},4p_{y},4p_{z}$) with the metal core (parameters
$U_{mm}<0$) and the ligand atoms cores (parameters $V_{ML}>0$).
Second term describes interaction of the ligand electrons with the
ligand cores (parameters $U_{ll}<0$), with the cores of the other
ligand atoms (parameters $V_{LL^{\prime}}>0$) and with the metal
core (parameter $V_{LM}>0$). Third and fourth terms describe the
resonance interactions in the ligand subsystem (parameters $\beta_{ml}<0$
and $\beta_{ll^{\prime}}<0$). Last term describes the Coulomb interactions
between electrons ($(ll^{\prime}\mid l^{\prime\prime}l^{\prime\prime\prime})$
are the corresponding two-electron integrals).

\subsubsection{Fockian for the \emph{l}-system}

The calculation of the wave function of the \emph{l}-system is the
prerequisite 

Variation principle applied to the effective Hamiltonian with the
trial function of the form eq. (\ref{eq:WaveFunction}) leads to the
self-consistent system of equations:
\begin{equation}
\begin{array}{ccl}
H_{d}^{eff}\Phi_{n}^{d} & = & E_{n}^{d}\Phi_{n}^{d}\\
H_{l}^{eff}\Phi_{0}^{l} & = & E_{0}^{l}\Phi_{0}^{l}\\
H_{d}^{eff} & = & H_{d}+\left\langle \Phi_{0}^{l}\mid H_{c}+H_{rr}\mid\Phi_{0}^{l}\right\rangle ,\\
H_{l}^{eff} & = & H_{l}+\left\langle \Phi_{0}^{d}\mid H_{c}+H_{rr}\mid\Phi_{0}^{d}\right\rangle .
\end{array}\label{Hdeff}
\end{equation}
In the above system the effective Hamiltonian $H_{d}^{eff}$ for the
$d$-electron subsystem depends on the wave function of the ligand
subsystem $\Phi_{0}^{l}$, and in its turn the effective Hamiltonian
$H_{l}^{eff}$ for the ligand susbsytem depends on the $d$-electrons'
wave functions $\Phi_{0}^{d}$. These equations must be solved self-consistently
as well. In the EHCF method Ref. \cite{Tch017} the Slater determinant
$\Phi_{0}^{l}$, is constructed of MO's of the $l$-system, obtained
from the Hartree-Fock equations in the INDO approximation for the
valence electrons of the ligands. In this case the transition from
the bare Hamiltonian $H_{l}^{eff}$ for the $l$-system to the corresponding
effective (dressed) Hamiltonian reduces to renormalization of one-electron
parameters related to the TMI: 
\begin{equation}
\begin{array}{c}
U_{ii}^{eff}=U_{ii}+\frac{1}{5}n_{d}\sum\limits _{\mu}g_{\mu i},\\
Z_{M}^{eff}=Z_{M}-n_{d},
\end{array}
\end{equation}
where $U_{ii}$ is the parameter of the interaction of $4s$ and $4p$-electrons
($i=4s$, $4p_{x}$, $4p_{y}$, $4p_{z}$) with the TMI core, $Z_{M}$
is TMI core charge, $g_{\mu i}=(\mu\mu|ii)-\frac{1}{2}(\mu i|i\mu)$
are the parameters of intraatomic Coulomb interactions. The $\Phi_{0}^{l}$
function thus obtained is used further for constructing the effective
Hamiltonian for the $d$-shell.

\subsection{Intersubsystem interaction operators\label{sub:Intersubsystem-interaction-opera}}

Here we introduce explicit definitions for the interaction operators
acting between the \emph{d}- and \emph{l}-system of a PTMC.

\subsubsection{\emph{dl}-Resonance operator\label{sub:dl-Resonance-operator}}

The resonance operator $H_{r}$ describing one-electron hopping between
the \emph{d}-shells and the ligands has the form: 
\begin{eqnarray}
H_{r} & = & H_{r}^{(+)}+H_{r}^{(-)},\nonumber \\
H_{r}^{(+)} & = & -\sum_{\sigma}\sum_{\mu,i}\sum_{\lambda}\beta_{\mu\lambda}(i)d_{\mu\sigma}^{+}(i)a_{\lambda\sigma},\label{ResonanceOperator}\\
H_{r}^{(-)} & = & -\sum_{\sigma}\sum_{\mu,i}\sum_{\lambda}\beta_{\mu\lambda}(i)a_{\lambda\sigma}^{+}d_{\mu\sigma}(i),\nonumber 
\end{eqnarray}
where $\lambda$ runs over $l$-MOs, $\beta_{\mu\lambda}(i)$ is the
resonance (hopping) integral between the $\mu$-th $d$-AO of the
$i$-th TMI and the $\lambda$-th $l$-MO, the fermion operator $d_{\mu\sigma}^{+}(i)$
creates an electron with the spin projection $\sigma$ on the $\mu$-th
$d$-AO of the $i$-th TMI, and $a_{\lambda\sigma}^{+}$ creates an
electron with the spin projection $\sigma$ on the $\lambda$-th $l$-MO.
One can easily check that the terms $H_{r}^{(\pm)}$ are the hermitean
conjugates of each other. Action of $H_{r}^{(+)}$apparently results
in positive ionization of the \emph{l}-system ($l\rightarrow d$ one-electron
transfer), that of $H_{r}^{(-)}$ refers to the negative ionization
of the \emph{l}-system ($d\rightarrow l$ one-electron transfer).
The above definition can be somewhat simplified by using the spinor
notation: 
\begin{eqnarray}
H_{r}^{(+)} & = & -\sum_{\mu,i}\sum_{\lambda}\beta_{\mu\lambda}(i)d_{\mu}^{+}(i)\hat{\iota}a_{\lambda},\label{eq:ResonanceOperatorSpinor}\\
H_{r}^{(-)} & = & -\sum_{\mu,i}\sum_{\lambda}\beta_{\mu\lambda}(i)a_{\lambda}^{+}\hat{\iota}d_{\mu}(i).\nonumber 
\end{eqnarray}

\subsubsection{\emph{dl}-Coulomb and exchange operators}

Although fundamentally we rely upon the INDO approximation for the
bare Hamiltonian for the PTMC we have to make certain concessions
and regrouping of terms in order to profit from the symmetries characteristic
for the atomic problem. Specifically we separate the Coulomb interaction
into symmetric (superscript {}``(s)'') and asymmetric (superscript
{}``(a)'') parts of which the first incures only the uniform shift
of the \emph{d}-levels in each given TMI whereas the second induces
the splitting of the otherwise degenerate \emph{d}-levels. The symmetry
mean by the symmetric part is that of the $SO(5)$ - i.e. of an arbitarary
orthogonal transformation of the \emph{d}-orbitals. These two contributions
further subdivide into an interaatomic (marked by subscript <<1>>)
part describing the interactions with the electrons in the noncorrelated
\emph{sp}-AOs of a given TMI and the interatomic\emph{ }one describing
the interactions with electrons residing on other atoms (marked by
subscript <<2>>). With these assumptions the Coulomb interaction
between the subsystems of a PTMC reads:

\begin{eqnarray}
H_{c} & = & H_{c}^{(s)}+H_{c}^{(a)}\nonumber \\
H_{c1}^{(s)} & = & \sum_{i}\sum_{L}\hat{n}_{d}(i)\left[\hat{n}_{s}(i)F_{i}^{0}(sd)+\hat{n}_{p}(i)F_{i}^{0}(pd)\right]\nonumber \\
H_{c2}^{(s)} & = & e^{2}\sum_{i}\sum_{L}\hat{n}_{d}(i)\hat{n}_{L}F_{0}(R_{Li})\\
H_{c1}^{(a)} & = & \sum_{i\mu}\sum_{L}\hat{n}_{\mu}(i)\hat{n}_{\gamma}(i)g_{\mu\gamma}^{pd}(i)\nonumber \\
H_{c2}^{(a)} & = & \sum_{mm^{\prime}}\hat{V}_{mm^{\prime}};\hat{n}_{d}(i)=\sum_{\mu}\hat{n}_{\mu}(i)\nonumber 
\end{eqnarray}
where $\hat{n}_{d}(i)$ and $\hat{n}_{l}$ are the operators of the
numbers of electrons in the $i$-th \emph{d}-shell and on the $l$-th
$l$-AO; $\hat{n}_{\mu}(i)$ is that for the $\mu$-th $d$-AO of
the $i$-th TMI; $L$ stands for all orbitals $l$ centered on the
ligand atom $L$; $\gamma=x,y,z$ $g_{\mu l}(i)$ are the energy parameters
(in fact -- the average Coulomb interaction integrals) and the interatomic
contributions $\hat{V}_{mm^{\prime}}$ are given by: 
\begin{eqnarray*}
\hat{V}_{mm^{\prime}} & = & \sum_{L}\sum_{\substack{l\in L\\
\sigma,\tau
}
}V_{mm^{\prime}}^{L}d_{m\sigma}^{+}d_{m^{\prime}\sigma}l_{\tau}^{+}l_{\tau}
\end{eqnarray*}
with 
\begin{equation}
V_{mm^{\prime}}^{L}=5\sqrt{4\pi}\sum_{k=2,4}\frac{(-1)^{m}}{\sqrt{2k+1}}\left(\begin{array}{ccc}
2 & k & 2\\
0 & 0 & 0
\end{array}\right)\left(\begin{array}{ccc}
2 & k & 2\\
m & m^{\prime}-m & -m^{\prime}
\end{array}\right)F_{k}(R_{L})Y_{k}^{m-m^{\prime}}(\theta_{L},\phi_{L}).\label{IonicCrystalField}
\end{equation}
where $(R_{L},\theta_{L},\phi_{L})$ are the spherical coordinates
of the ligand atom $L$ (relative to a TMI consequently located in
the center of the coordinate frame); $Y_{k}^{m-m^{\prime}}(\theta_{L},\phi_{L})$
are the spherical functions with the phases defined following Condon
and Shortley.\cite{CondonShortley} Functions $F_{k}(R_{L})$ are
the integrals of squares of the radial parts $R_{nl}(r)$ of the atomic
$d$-functions:

\begin{equation}
F_{k}(R)=R^{-(k+1)}\int\limits _{0}^{R}r^{k}R_{nl}^{2}(r)r^{2}dr+R^{k}\int\limits _{R}^{\infty}r^{-(k+1)}R_{nl}^{2}(r)r^{2}dr\label{FFunction}
\end{equation}
depend on the distance $R_{L}$ from the atom of metal to the atom
$L$. For the Slater AOs $R_{nl}(r)$ the functions $F_{k}(R)$ are
explicitly known.\cite{Bersukerbook}

Experience of usage of the EHCF numerical procedure acquired so far
shows that the effect of the Coulomb splitting (asymmetric part of
the Coulomb interaction) can be safely neglected when it goes about
estimating the energy denominators in the resolvent in Section \ref{sub:Resolvents-for-the-Configuration}
which is done consistently.

The exchange operator decomposes analogously with that difference
that in the INDO approximation only the intraatomic contribution on
the TMIs is present. 
\begin{eqnarray}
H_{x} & = & H_{x}^{(s)}+H_{x}^{(a)}
\end{eqnarray}
We however restrict ourselves by the symmetric part of the intraatomic
intershell exchange for the time being: 
\begin{eqnarray}
H_{x}^{(s)} & = & -2\sum_{i}\sum_{\substack{\mu\in d\text{-}\mathrm{AO}\\
l\in l\text{-}\mathrm{AO}
}
}\hat{S}_{d}(i)\left[K_{sd}(i)\hat{S}_{s}(i)+K_{pd}(i)\hat{S}_{p}(i)\right]\label{eq:IntraAtomicExchange}\\
\hat{S}_{d}(i) & = & \sum_{\mu}\hat{S}_{\mu}(i)\nonumber 
\end{eqnarray}
where $\hat{S}_{\mu}(i)$ and $\hat{S}_{l}(i)$ are the electron spin
operators for electrons in the $\mu$-th $d$-AO of the $i$-th TMI
and the $l$-th $l$-AO of the $i$-th TMI. The exchange parameters
are given by: 
\begin{eqnarray}
K_{sd}(i) & = & \frac{1}{5}G_{sd}^{2}(i)\nonumber \\
K_{pd}(i) & = & \frac{2}{15}G_{pd}^{1}(i)+\frac{113}{1225}G_{pd}^{3}(i)\label{eq:IntraatomicExchangeParameters}
\end{eqnarray}

\section{Expansion of the resolvent eq. (\ref{eq:ResolventItalic})\label{sec:Expansion-of-the} }

Inserting the expansion of the resolvent eq. (\ref{eq:ResolventItalic})
in eq. (\ref{eq:EffectiveHamiltonian}) we obtain: 
\begin{eqnarray}
H^{\mathrm{eff}}(E) & = & PH_{0}P+PH_{x}P+\nonumber \\
 &  & +PH_{r}QR_{0}QH_{r}P+\nonumber \\
 &  & +PH_{r}QR_{0}QH_{x}QR_{0}QH_{r}P+\label{eq:EffectiveExpansion}\\
 &  & +PH_{r}QR_{0}QH_{r}QR_{0}QH_{r}QR_{0}QH_{r}P+\nonumber \\
 &  & +PH_{r}QR_{0}QH_{x}QR_{0}QH_{x}QR_{0}QH_{r}P+\ldots\nonumber 
\end{eqnarray}
In the above expression (and in its continuation whatsoever) the odd
powers of $H_{r}$ disappear since the operator $H_{r}$ changes the
number of electrons in the $d$- and\emph{ l}-systems by one in either
sum, and in case when the number of such changes is odd the total
number of electrons in the $d$- and/or \emph{l}-systems cannot be
conserved. Thus the number of the $H_{r}$ multipliers must be even.

Further analysis can be based on the notion of \textquotedbl{}graduality\textquotedbl{}
of the complementary projection operator $Q$ and the resolvent expressed
by the following expansion for these quantities: 
\begin{equation}
\begin{array}{c}
Q=\bigoplus\limits _{n}Q^{(n\pm)}\\
R_{0}=\bigoplus\limits _{n}R^{(n\pm)}
\end{array}\label{Graduality}
\end{equation}
where $n$ refers to the number of electrons added to (\textquotedblright{}$+$\textquotedblright{})
or taken from (\textquotedblright{}$-$\textquotedblright{}) the $d$-system
(this notation precisely refers to the degree of ionization of the
\emph{l}-system), and $\oplus$ stands for the direct (block) sum
of the corresponding matrices. For the terms in eq. (\ref{Graduality})
the following orthogonality conditions hold: 
\[
\begin{array}{c}
Q^{(n\pm)}R^{(m\pm)}Q^{(p\pm)}=\delta_{nm}\delta_{mp}R^{(m\pm)}\\
Q^{(m\pm)}R^{(n\pm)}P=PR^{(n\pm)}Q^{(m\pm)}=0
\end{array}
\]

With this decomposition for the resolvent and graduation of the $\mathrm{Im}Q$
configuration subspace the effective Hamiltonian rewrites: 
\begin{eqnarray}
H^{\mathrm{eff}}(E) & = & P\left(H_{0}+H_{x}\right)P+\nonumber \\
 & + & PH_{r}Q^{(1\pm)}R^{(1\pm)}Q^{(1\pm)}H_{r}P+\nonumber \\
 & + & PH_{r}Q^{(1\pm)}R^{(1\pm)}Q^{(1\pm)}H_{x}Q^{(1\pm)}R^{(1\pm)}Q^{(1\pm)}H_{r}P+\nonumber \\
 & + & PH_{r}Q^{(1\pm)}R^{(1\pm)}Q^{(1\pm)}H_{r}Q^{(2\pm)}R^{(2\pm)}Q^{(2\pm)}H_{r}Q^{(1\pm)}R^{(1\pm)}Q^{(1\pm)}H_{r}P+\label{eq:EffectiveExpansionLong}\\
 & + & PH_{r}QR_{0}QH_{x}QR_{0}QH_{x}QR_{0}QH_{r}P+...\nonumber 
\end{eqnarray}
where the terms with <<+>> and <<->> superscripts are summed separately.

As for the terms containing the powers of the exchange operator $H_{x}QR_{0}$
only they seem to be summable since they involve only the terms of
the same graduality. It looks like:

\begin{eqnarray*}
 &  & PH_{r}QR_{0}QH_{x}QR_{0}QH_{r}P+\\
 &  & PH_{r}QR_{0}QH_{x}QR_{0}QH_{x}QR_{0}QH_{r}P+\\
 &  & PH_{r}QR_{0}QH_{x}QR_{0}QH_{x}QR_{0}QH_{x}QR_{0}QH_{r}P+...\\
 & = & PH_{r}QR_{0}\sum_{m=1}^{\infty}\left(QH_{x}QR_{0}\right)^{m}QH_{r}P\\
 & = & PH_{r}QR_{0}\frac{QH_{x}QR_{0}}{Q-R_{0}QH_{x}Q}QH_{r}P=\\
 & = & PH_{r}QR_{0}\frac{QH_{x}Q}{R_{0}^{-1}-QH_{x}Q}QH_{r}P
\end{eqnarray*}
which represents nothing but multiple scattering of an electron wandering
in the l-system by all possible TMIs in all possible combinations.
For the time being we retain only the first order term in $H_{x}$.
Thus we obtain: 
\begin{eqnarray}
H^{\mathrm{eff}}(E) & \approx & PH_{0}P+PH_{x}P+\nonumber \\
 & + & PH_{r}QR_{0}QH_{r}P+\nonumber \\
 & + & PH_{r}QR_{0}QH_{x}QR_{0}QH_{r}P+\label{eq:EffectiveHamiltonianCompleteExpansion}\\
 & + & PH_{r}Q^{(1+)}R^{(1+)}Q^{(1+)}H_{r}Q^{(2+)}R^{(2+)}Q^{(2+)}H_{r}Q^{(1+)}R^{(1+)}Q^{(1+)}H_{r}P+\nonumber \\
 & + & PH_{r}Q^{(1-)}R^{(1-)}Q^{(1-)}H_{r}Q^{(2-)}R^{(2-)}Q^{(2-)}H_{r}Q^{(1-)}R^{(1-)}Q^{(1-)}H_{r}P\nonumber 
\end{eqnarray}
Since the model configuration subspace is that of the nonionized \emph{l}-system
the resonance operator eq. (\ref{ResonanceOperator}) can enter in
the answer in even powers. We classify the terms in the effective
interaction stemming from the partition procedure eq. (\ref{eq:EffectiveHamiltonian})
according to the degree of the ionization these terms introduce to
the \emph{l}-system. In the lowest order we can write:

\begin{equation}
H_{rr}=H_{r}R_{0}H_{r}=H_{rr}^{(0)}+H_{rr}^{(2+)}+H_{rr}^{(2-)}\label{eq:SecondPowerOfHopping}
\end{equation}
The ionization components of the operator $H_{rr}$ have the form:

\begin{equation}
\begin{array}{c}
H_{rr}^{(2+)}=\sum\limits _{\mu i\lambda\sigma}\sum\limits _{\nu j\kappa\tau}\beta_{\mu\lambda}(i)\beta_{\nu\kappa}(j)d_{\mu\sigma}^{+}(i)a_{\lambda\sigma}R^{(1+)}d_{\nu\tau}^{+}(j)a_{\kappa\tau}\\
H_{rr}^{(2-)}=\sum\limits _{\mu i\lambda\sigma}\sum\limits _{\nu j\kappa\tau}\beta_{\mu\lambda}(i)\beta_{\nu\kappa}(j)d_{\mu\sigma}(i)a_{\lambda\sigma}^{+}R^{(1-)}d_{\nu\tau}(j)a_{\kappa\tau}^{+}\\
H_{rr}^{(0)}=\sum\limits _{\mu i\lambda\sigma}\sum\limits _{\nu j\kappa\tau}\beta_{\mu\lambda}(i)\beta_{\nu\kappa}(j)\times\\
\left\{ a_{\lambda\sigma}^{+}d_{\mu\sigma}(i)R^{(1+)}d_{\nu\tau}^{+}(j)a_{\kappa\tau}\right.+\\
+\left.d_{\mu\sigma}^{+}(i)a_{\lambda\sigma}R^{(1-)}a_{\kappa\tau}^{+}d_{\nu\tau}(j)\right\} 
\end{array}\label{eq:EffectiveHoppingComponents}
\end{equation}
Comparing eq. (\ref{eq:EffectiveExpansionLong}) with the definitions
eqs. (\ref{eq:SecondPowerOfHopping}), (\ref{eq:EffectiveHoppingComponents})
we can identify the products $H_{r}Q^{(1\pm)}R^{(1\pm)}Q^{(1\pm)}H_{r}$
with components $H_{rr}^{(2\pm)}$ so that eq. (\ref{eq:EffectiveHamiltonianCompleteExpansion})
rewrites:

\begin{eqnarray}
H^{\mathrm{eff}}(E) & \approx & PH_{0}P+PH_{x}P+PH_{rr}^{(0)}P+\nonumber \\
 & + & PH_{r}^{(+)}Q^{(1+)}R_{0}^{(1+)}Q^{(1+)}H_{x}Q^{(1+)}R_{0}^{(1+)}Q^{(1+)}H_{r}^{(+)}P+\nonumber \\
 & + & PH_{r}^{(-)}Q^{(1-)}R_{0}^{(1-)}Q^{(1-)}H_{x}Q^{(1-)}R_{0}^{(1-)}Q^{(1-)}H_{r}^{(-)}P+\\
 & + & PH_{rr}^{(2+)}Q^{(2+)}R^{(2+)}Q^{(2+)}H_{rr}^{(2+)}P+\nonumber \\
 & + & PH_{rr}^{(2-)}Q^{(2-)}R^{(2-)}Q^{(2-)}H_{rr}^{(2-)}P\nonumber 
\end{eqnarray}
This represents the approximate effective Hamiltonian acting in the
subspace with the fixed distribution of electrons between the \emph{d}-
and \emph{l}-systems.

\section{Resolvents and energy estimates in the outer subspaces\label{sec:Resolvents}}

\subsection{Resolvents for the configuration space partition\label{sub:Resolvents-for-the-Configuration}}

The components for the resolvent acting in the configuration subspace
$\mathrm{Im}Q$ for the subspace with the singly ionized ligands have
the form: 
\[
\begin{array}{c}
{\displaystyle R^{(1-)}=\sum\limits _{k}\mathcal{\mathbb{O}}_{-}^{k}\otimes\left(\sum\limits _{\kappa\tau}D^{(1-)}(k\kappa)a_{\kappa\tau}\left\vert \Phi_{L}\right\rangle \left\langle \Phi_{L}\right\vert a_{\kappa\tau}^{+}\right)}\\
{\displaystyle R^{(1+)}=\sum\limits _{k}\mathcal{\mathbb{O}}_{+}^{k}\otimes\left(\sum\limits _{\kappa\tau}D^{(1+)}(k\kappa)a_{\kappa\tau}^{+}\left\vert \Phi_{L}\right\rangle \left\langle \Phi_{L}\right\vert a_{\kappa\tau}\right)}
\end{array}
\]
(we remind that the superscripts for the resolvents refer to the ionization
degree of the \emph{l}-system). They involve the projection operators
in the \emph{d}-system 
\begin{eqnarray}
\mathbb{O} & = & \mathbb{O}(\left\{ n_{d}(j)\right\} );\nonumber \\
\mathbb{O}_{+}^{k} & = & \mathcal{\mathbb{O}}(\left\{ n_{d}(j)\right\} _{j\neq k},n_{d}(k)+1);\label{eq:d-projector}\\
\mathbb{O}_{-}^{k} & = & \mathcal{\mathbb{O}}(\left\{ n_{d}(k)\right\} _{j\neq k},n_{d}(k)-1)\nonumber 
\end{eqnarray}
are those projecting to the subspaces with the respective fixed numbers
of electrons in the $d$-shell of the $j$-th TMI in the polynuclear
complex. The projection operators further subdivide into components
corresponding to the resulting total spin of the affected \emph{d}-shell,
that is: 
\[
\mathbb{O}_{\pm}^{k}=\mathbb{O}_{\pm}^{k}\left(S(k)+\frac{1}{2}\right)+\mathbb{O}_{\pm}^{k}\left(S(k)-\frac{1}{2}\right).
\]
 Due to cumbersome notation we shall not always indicate this subdivision,
but shall keep track of it since it is necessary (see Sections \ref{sec:Effective-Hamiltonian-contributi}
and \ref{sec:Specific-contributions-to}).

The projection and Fermi operators obey the following commutation
relations:\cite{Tch017} 
\begin{equation}
\begin{array}{c}
d(j)\mathcal{\mathbb{O}}_{+}^{j}=\mathcal{\mathbb{O}}d(j);\ d^{+}(j)\mathbb{O}_{-}^{j}=\mathcal{\mathbb{O}}d^{+}(j)\\
\mathbb{O}_{+}^{j}d^{+}(j)=d^{+}(j)\mathbb{O};\ \mathbb{O}_{-}^{j}d(j)=d(j)\mathcal{\mathbb{O}}
\end{array}\label{eq:d-ProjectorCommutation}
\end{equation}
The energy denominators: 
\[
\begin{array}{c}
D^{(1+)}(k\kappa)=\left[E-(I_{\kappa}-A(k)-g_{\kappa}(k))\right]{}^{-1}\\
D^{(1-)}(k\kappa)=\left[E-(I(k)-A_{\kappa}-g_{\kappa}(k))\right]{}^{-1}
\end{array}
\]
where $g_{\kappa}(k)$ is the Coulomb interaction between an electron
and a hole in the $k$-th $d$-shell and $\kappa$-th $l$-MO and
the operators and \emph{I}'s and \emph{A}'s are the ionization potentals
(IPs) and electronic affinities (EAs) of the ligands and \emph{d}-shells.
Here we as well will be keeping track of the spins of the \emph{k}-th
\emph{d}-shell resulting from the electron addition or abstraction.

The estimates for the IPs and EAs for noncorrelated \emph{l}-system
MOs are taken according to the Koopmans theorem: 
\begin{equation}
I_{\kappa},A_{\kappa}=-\varepsilon_{\kappa}\label{eq:KoopmansTheorem}
\end{equation}
coming from the semiempirical SCF procedure as applied to the \emph{l}-system.
The ionization potentials and electron affinities for the \emph{d}-shells
($I(k),A(k)$) are more complicated since they have to account for
the difference between the electron addition and electron subtraction
to/from the \emph{d}-shell as well as for the total spin of the \emph{d}-shell
emerging as a result of either of the processes. Using the expression
for the bare energies eq. (\ref{eq:d-shellBareEnergy}) for the relevant
states we obtain: 
\begin{eqnarray*}
I(k) & = & -U_{dd}(k)-g_{dd}(k)\left(n_{d}(k)-1\right)+\begin{cases}
-K_{dd}(k)\left(S_{d}(k)+\frac{3}{4}\right) & \mathrm{increasing\, spin}\\
K_{dd}(k)\left(S_{d}(k)+\frac{1}{4}\right) & \mathrm{decreasing\, spin}
\end{cases}\\
A(k) & = & -U_{dd}(k)-g_{dd}(k)n_{d}(k)+\begin{cases}
K_{dd}(k)\left(S_{d}(k)+\frac{3}{4}\right) & \mathrm{increasing\, spin}\\
-K_{dd}(k)\left(S_{d}(k)+\frac{1}{4}\right) & \mathrm{decreasing\, spin}
\end{cases}
\end{eqnarray*}
The difference between the IP's and EA's for the different resulting
spins yields in both cases precisely $\pm K_{dd}(k)\left(2S_{d}(k)+1\right)$
which with attention to the notation coincides with the result given
in Ref. \cite{WeiheGuedel1997}. By this the resolvent relevant for
the partition eq. (\ref{eq:EffectiveHamiltonian}) is defined.

\subsection{Resolvents for the magnetic limit\label{sub:Resolvents-for-the-magnetic}}

The resolvent appearing in the partition eq. (\ref{eq:HamiltonianMagneticSubspace})
is the most peculiar one since as one can see from eq. (\ref{eq:DressedEffectiveHopping})
of Section \ref{sec:Effective-Hamiltonian-contributi} it describes
the contribution of a process which can be called a dressed effective
hopping and the relevant excited states in the configuration subspace
$\mathrm{Im}\mathfrak{Q}$ involve the states with distribution of
\emph{d}-electrons different from the fixed one as well as the excitations
of the \emph{l}-system. The operators projecting on the basis states
in $\mathrm{Im}\mathfrak{Q}$ are thus: 
\begin{eqnarray*}
\mathcal{\mathbb{O}}_{+}^{j}\mathcal{\mathbb{O}}_{-}^{i} & \otimes & \left\vert \Phi_{\kappa\rightarrow\lambda}^{S}\right\rangle \left\langle \Phi_{\kappa\rightarrow\lambda}^{S}\right\vert \\
\mathcal{\mathbb{O}}_{+}^{j}\mathcal{\mathbb{O}}_{-}^{i} & \otimes & \left\vert \Phi_{\kappa\rightarrow\lambda}^{T\gamma}\right\rangle \left\langle \Phi_{\kappa\rightarrow\lambda}^{T\gamma}\right\vert 
\end{eqnarray*}
which are to be supplied with the respective energy denominators:
\begin{eqnarray}
\mathfrak{D}_{ijS\kappa\rightarrow\lambda}^{(0)} & = & \left[\mathfrak{E}-\left(I(i)-A(j)-g_{ij}+E_{R}^{(i\rightarrow j)}\right)\right.\nonumber \\
 & - & \left.\left(1-\delta_{\kappa\lambda}\right)\left(I_{\kappa}^{(i\rightarrow j)}-A_{\lambda}^{(i\rightarrow j)}-g_{\lambda\kappa}+2K_{\lambda\kappa}\right)\right]^{-1}\nonumber \\
\mathfrak{D}_{ijT\kappa\rightarrow\lambda}^{(0)} & = & \left[\mathfrak{E}-\left(I(i)-A(j)-g_{ij}+E_{R}^{(i\rightarrow j)}\right)\right.\label{eq:MagneticResolventDenominator}\\
 & - & \left.\left(1-\delta_{\kappa\lambda}\right)\left(I_{\kappa}^{(i\rightarrow j)}-A_{\lambda}^{(i\rightarrow j)}-g_{\lambda\kappa}\right)\right]^{-1}\nonumber 
\end{eqnarray}
where $I(i)$ and $A(j)$ are the IP and EA for the \emph{i}-th and
\emph{j}-th \emph{d}-shells respectively as defined in Section \ref{sub:Resolvents-for-the-Configuration}.
They contain the dependence of the energy of the MMCT states on the
spins of the \emph{d}-shells involved in the transfer. The quantity
$g_{ij}$ is the average Coulomb interaction parameter for electron
and hole residing in the respective \emph{d}-shells, which can be
taken in a simple form; $g_{\lambda\kappa}$ and $K_{\lambda\kappa}$
are, respectively, the Coulomb and exchange integrals for the pair
of $l$-MOs coming from the semiempirical SCF calculation; $E_{R}^{(i\rightarrow j)}$
is the reorganization energy of the \emph{l}-system acquired under
the transition, which can be estimated as: 
\[
E_{R}^{(i\rightarrow j)}=\sum_{L}q_{L}(g_{iL}-g_{jL})
\]
where the effective charges $q_{L}$ in the \emph{l}-system come from
the semiempirical SCF calculation for the \emph{l}-system in the electrostatic
field induced by the fixed distribution of \emph{d}-electrons. The
ionization potentials and electron affinities $I_{\kappa}^{(i\rightarrow j)}$
and $A_{\lambda}^{(i\rightarrow j)}$ are evaluated according to the
Koopmans theorem as the orbital energies of the corresponding MOs
shifted by the first order correction from the electrostatic field
induced by the charge transferred between the \emph{d}-shells. That
is: 
\[
I_{\kappa}^{(i\rightarrow j)}=I_{\kappa}+\sum_{L}(g_{iL}-g_{jL})\sum_{l\in L}\left|\left\langle Ll\mid\kappa\right\rangle \right|^{2}
\]
Further energy denominators required for taking into account the states
in the configuration subspace $\mathrm{Im}Q^{(2\pm)}\mathfrak{Q}$
are those corresponding to double ionization of the \emph{l}- and
\emph{d}-systems: 
\begin{eqnarray*}
D_{ijS\kappa\lambda}^{(2-)} & = & \left[E-\left(I(i)+I(j)-A_{\kappa}-A_{\lambda}+g_{ij}+g_{\kappa\lambda}+E_{R}^{(i\rightarrow j)}\right)\right.\\
 & - & \left.\left(1-\delta_{\kappa\lambda}\right)\left(I_{\kappa}^{(i\rightarrow j)}-A_{\lambda}^{(i\rightarrow j)}-g_{\lambda\kappa}+2K_{\lambda\kappa}\right)\right]^{-1}\\
D_{ijS\kappa\lambda}^{(2+)} & = & \left[E-\left(I_{\kappa}+I_{\lambda}-A(i)-A(j)+g_{ij}+g_{\kappa\lambda}+E_{R}^{(i\rightarrow j)}\right)\right.\\
 & - & \left.\left(1-\delta_{\kappa\lambda}\right)\left(I_{\kappa}^{(i\rightarrow j)}-A_{\lambda}^{(i\rightarrow j)}-g_{\lambda\kappa}+2K_{\lambda\kappa}\right)\right]^{-1}
\end{eqnarray*}
The projection operators to the configurations in the doubly ionized
subspace are:

\begin{eqnarray*}
\mathcal{\mathbb{O}}_{+}^{j}\mathcal{\mathbb{O}}_{+}^{i} & \otimes & \left\vert \Phi_{\kappa\rightarrow\lambda}^{S(2+)}\right\rangle \left\langle \Phi_{\kappa\rightarrow\lambda}^{S(2+)}\right\vert \\
\mathcal{\mathbb{O}}_{-}^{j}\mathcal{\mathbb{O}}_{-}^{i} & \otimes & \left\vert \Phi_{\kappa\rightarrow\lambda}^{T\gamma(2-)}\right\rangle \left\langle \Phi_{\kappa\rightarrow\lambda}^{T\gamma(2-)}\right\vert 
\end{eqnarray*}

\subsection{Resolvents for the \emph{l}-system polarization}

For the \emph{l}-system only singly excited states over its singlet
(closed shell) ground state are previewed.\cite{DeLothCassouxDaudeyMalrieu,CharlotVerdaguerJournauxDeLothDaudey,deLothDaudeyAstheimerWalzHaase}
These states apparently classify according to their total spin which
can be thus singlet and triplet. For that reason the resolvent eq.
(\ref{eq:PolarizationResolvent}) decomposes 
\begin{equation}
\begin{array}{c}
\mathcal{R}(\Omega)=\mathcal{R}_{S}(\Omega)+\mathcal{R}_{T}(\Omega)\\
\mathcal{R}_{S}^{(0)}(\Omega)=\sum_{\substack{\lambda\kappa}
}\left\vert \Phi_{\kappa\rightarrow\lambda}^{S}\right\rangle \mathcal{D}_{S\kappa\rightarrow\lambda}^{(0)}\left\langle \Phi_{\kappa\rightarrow\lambda}^{S}\right\vert ,\\
\mathcal{R}_{T}^{(0)}(\Omega)=\sum_{\substack{\lambda\kappa\\
\gamma
}
}\left\vert \Phi_{\kappa\rightarrow\lambda}^{T\gamma}\right\rangle \mathcal{D}_{T\kappa\rightarrow\lambda}^{(0)}\left\langle \Phi_{\kappa\rightarrow\lambda}^{T\gamma}\right\vert ,
\end{array}\label{eq:l-SystemResolventDecomposition}
\end{equation}
where the superscript on the left side refers to the fact that the
\emph{l}-system is not ionized in this configuration subspace; $\left\vert \Phi_{\kappa\rightarrow\lambda}^{S}\right\rangle $
and $\left\vert \Phi_{\kappa\rightarrow\lambda}^{T\gamma}\right\rangle $
stand respectively for the singlet and triplet excited states of the
\emph{l}-system with one electron excited from the MO $\left|\kappa\right\rangle $
occupied in the ground state to the MO $\left|\lambda\right\rangle $
empty in the ground state and $\gamma$ denotes the component ($x,y,z$)
of the triplet state. The excited states of the $l$-system yield
the energy denominators 
\begin{eqnarray*}
\mathcal{D}_{S\kappa\rightarrow\lambda}^{(0)} & = & \left[\Omega-\left(I_{\kappa}-A_{\lambda}-g_{\lambda\kappa}+2K_{\lambda\kappa}\right)\right]^{-1}\\
\mathcal{D}_{T\kappa\rightarrow\lambda}^{(0)} & = & \left[\Omega-\left(I_{\kappa}-A_{\lambda}-g_{\lambda\kappa}\right)\right]^{-1}
\end{eqnarray*}
naturally independent on the projections in the triplet state. Here
$I_{\kappa}$ and $A_{\lambda}$ are respectively ionization potential
and electron affinity of the \emph{l}-system as defined by eq. (\ref{eq:KoopmansTheorem}),
$g_{\lambda\kappa}$ is the Coulomb integral for the pair of $l$-MOs,
$K_{\lambda\kappa}$ is the exchange integral for the pair of $l$-MOs.

\section{Contributions to the effective Hamiltonian\label{sec:Effective-Hamiltonian-contributi}}

The contributions to the effective Hamiltonian for the \emph{d}-system
as described in Section \ref{sec:Specific-contributions-to} are the
averages over the ground state of the \emph{l}-system $\left\vert \Phi_{L}\right\rangle $.
The averages of the diagonal (symmetric) parts of the interaction
operators comes out trivially. Nontrivial contributions come from
the operators capable to excite the \emph{l}-system. We consider them
one by one in the following Subsections.

\subsection{Effective one-electron hopping between TMIs and related terms\label{sub:Effective-one-electron-hopping}}

\subsubsection{Preliminaries}

We start with the action of the operator $H_{rr}^{(0)}$ on the ground
state of the $l$-system. It yields: 
\[
\begin{array}{c}
H_{rr}^{(0)}\left\vert \Phi_{L}\right\rangle =\sum\limits _{\mu i\lambda\sigma}\sum\limits _{\nu j\kappa\tau}\beta_{\mu\lambda}(i)\beta_{\nu\kappa}(j)\times\\
\times\left\{ a_{\lambda\sigma}^{+}d_{\mu\sigma}(i)\sum\limits _{k\kappa^{\prime}\tau^{\prime}}\mathcal{\mathbb{O}}_{+}^{k}D^{(1+)}(k\kappa^{\prime})a_{\kappa^{\prime}\tau^{\prime}}\left\vert \Phi_{L}\right\rangle \left\langle \Phi_{L}\right\vert a_{\kappa^{\prime}\tau^{\prime}}^{+}d_{\nu\tau}^{+}(j)a_{\kappa\tau}\left\vert \Phi_{L}\right\rangle \right.\\
+\left.d_{\mu\sigma}^{+}(i)a_{\lambda\sigma}\sum\limits _{k\kappa^{\prime}\tau^{\prime}}\mathcal{\mathbb{O}}_{-}^{k}D^{(1-)}(k\kappa^{\prime})a_{\kappa^{\prime}\tau^{\prime}}^{+}\left\vert \Phi_{L}\right\rangle \left\langle \Phi_{L}\right\vert a_{\kappa^{\prime}\tau^{\prime}}a_{\kappa\tau}^{+}d_{\nu\tau}(j)\left\vert \Phi_{L}\right\rangle \right\} 
\end{array}
\]
Employing the anticommutation relations for the Fermi operators: 
\[
a_{i}a_{j}=-a_{j}a_{i};\ a_{i}^{+}a_{j}^{+}=-a_{j}^{+}a_{i}^{+};\ a_{i}^{+}a_{j}+a_{j}a_{i}^{+}=\delta_{ij}
\]
and the values of the averages of the products of the latter over
the single-determinant functions: 
\[
\left\langle \Phi_{L}\right\vert a_{\kappa^{\prime}\tau^{\prime}}^{+}a_{\kappa\tau}\left\vert \Phi_{L}\right\rangle =\delta_{\kappa\kappa^{\prime}}\delta_{\tau\tau^{\prime}}n_{\kappa\tau};\ \left\langle \Phi_{L}\right\vert a_{\kappa^{\prime}\tau^{\prime}}a_{\kappa\tau}^{+}\left\vert \Phi_{L}\right\rangle =\delta_{\kappa\kappa^{\prime}}\delta_{\tau\tau^{\prime}}(1-n_{\kappa\tau})
\]
we get: 
\[
\begin{array}{c}
H_{rr}^{(0)}\left\vert \Phi_{L}\right\rangle =-\sum\limits _{\mu i\lambda\sigma}\sum\limits _{\nu j\kappa\tau}\beta_{\mu\lambda}(i)\beta_{\nu\kappa}(j)\times\\
\left\{ \sum\limits _{k}\delta_{kj}a_{\lambda\sigma}^{+}d_{\mu\sigma}(i)a_{\kappa\tau}\left\vert \Phi_{L}\right\rangle D^{(1+)}(k\kappa)\mathbb{O}_{+}^{k}d_{\nu\tau}^{+}(j)n_{\kappa\tau}\right.+\\
-\left.\sum\limits _{k}\delta_{kj}d_{\mu\sigma}^{+}(i)a_{\lambda\sigma}a_{\kappa\tau}^{+}\left\vert \Phi_{L}\right\rangle D^{(1-)}(k\kappa)\mathbb{O}_{-}^{k}d_{\nu\tau}(j)(1-n_{\kappa\tau})\right\} 
\end{array}
\]
Complementing this by the projection operator $\mathbb{O}$ from the
left and performing its commutation with the Fermi operators according
to eq. (\ref{eq:d-ProjectorCommutation}) we get 
\begin{equation}
\begin{array}{c}
H_{rr}^{(0)}\left\vert \Phi_{L}\right\rangle =\sum\limits _{\mu i\lambda\sigma}\sum\limits _{\nu j\kappa\tau}\beta_{\mu\lambda}(i)\beta_{\nu\kappa}(j)\times\\
\left\{ \mathcal{\mathbb{O}}_{-}^{i}\mathcal{\mathbb{O}}_{+}^{j}d_{\mu\sigma}(i)d_{\nu\tau}^{+}(j)a_{\lambda\sigma}^{+}a_{\kappa\tau}\left\vert \Phi_{L}\right\rangle D^{(1+)}(j\kappa)n_{\kappa}\right.+\\
+\left.\mathcal{\mathbb{O}}_{+}^{i}\mathcal{\mathbb{O}}_{-}^{j}d_{\mu\sigma}^{+}(i)d_{\nu\tau}(j)a_{\lambda\sigma}a_{\kappa\tau}^{+}\left\vert \Phi_{L}\right\rangle D^{(1-)}(j\kappa)(1-n_{\kappa})\right\} 
\end{array}\label{eq:DressedEffectiveHopping}
\end{equation}
which describes the process which admixes the states in the configuration
subspace $\mathrm{Im}\mathcal{P}\mathfrak{P}P$ to those in the subspace
$\mathrm{Im}\mathcal{P}\mathfrak{Q}Q^{(1\pm)}$. It is a starting
point for further evaluations.

Due to the definition of the projection operators $\mathfrak{P}$
and $\mathfrak{Q}$ the Fermi operators referring to the \emph{l}-systems
commute with them so that they act directly on the wave function $\left|\Phi_{L}\right\rangle $
and produce the singlet $\left|\Phi_{\kappa\rightarrow\lambda}^{S}\right\rangle $
and three components of the triplet $\left|\Phi_{\kappa\rightarrow\lambda}^{T\gamma}\right\rangle $
excited states of the ligands. This shows that in the general case
a motion of an electron between the \emph{d}-shells is accompanied
by the charge and/or spin polarization of the ligands. It had been
checked that the excitation energies in the \emph{l}-system of the
exemplary molecules considered in the present paper (see Section \ref{sec:Calculations-results})
can be estimated of about 10 eV, thus their contribution is strongly
damped.

An important exception is represented by the terms in eq. (\ref{eq:DressedEffectiveHopping})
where $\kappa=\lambda$. In this case $\sigma=\tau$ also holds and
the \emph{l}-system incidentally acquires no excitation. This singles
out the following terms from eq. (\ref{eq:DressedEffectiveHopping}):
\begin{equation}
\begin{array}{c}
\sum\limits _{i\neq j\mu\nu}\sum\limits _{\lambda\tau}\beta_{\mu\lambda}(i)\beta_{\nu\lambda}(j)\times\\
\left\{ \mathcal{\mathbb{O}}_{-}^{i}\mathcal{\mathbb{O}}_{+}^{j}d_{\mu\tau}(i)d_{\nu\tau}^{+}(j)\left\vert \Phi_{L}\right\rangle D^{(1+)}(j\lambda)n_{\lambda}\right.+\\
+\left.\mathcal{\mathbb{O}}_{+}^{i}\mathcal{\mathbb{O}}_{-}^{j}d_{\mu\tau}^{+}(i)d_{\nu\tau}(j)\left\vert \Phi_{L}\right\rangle D^{(1-)}(j\lambda)(1-n_{\lambda})\right\} 
\end{array}\label{eq:DressedEffectiveHoppingLigandDiagonal}
\end{equation}
This expression generalizes and provides the numerical estimate to
the matrix elements responsible for the MMCT admixture.\cite{WeiheGuedel1996,WeiheGuedel1997,WeiheGuedelToftlund,WeihGuedel2000}

This is apparently the one-electron operator acting in the $d$-system
and describing electron transfers between the $d$-shells of different
TMIs through the $l$-system. Two groups of channels for such transfers
do exist: the occupied and empty MOs of the $l$-system. In the first
case (occupied $\kappa$-th $l$-MO) an electron is captured from
it by the $j$-th $d$-shell and then the hole thus emerged is healed
on account of an electron coming from the $i$-th $d$-shell. In the
second case (empty $\kappa$-th $l$-MO) an electron leaves the $j$-th
\emph{d}-shell and then lands in the $i$-th $d$-shell. Despite an
unsymmetric appearance this effective operator is hermitean. The terms
with $i=j$ were precisely the covalent contributions to the effective
crystal field.\cite{Tch017}

\subsubsection{Averaging over the \emph{l}-system\label{sub:Averaging-over-the}}

As in the original EHCF derivation\cite{Tch017} the transition from
the effective Hamiltonian for the (P)TMC acting in the configuration
subspace with the fixed distribution of electrons between the \emph{d}-
and \emph{l}-systems to that for the \emph{d}-system only is performed
by averaging the former over the ground state of the \emph{l}-system
represented by the single determinant wave function $\left|\Phi_{L}\right\rangle $.
In variance with the case of TMC in PTMC we additionally assumed that
the $\left|\Phi_{L}\right\rangle $ must be calculated for the fixed
distribution of the \emph{d}-electrons among several \emph{d}-shells
of a PTMC. Formally it is reflected in the requirement that the effective
Hamiltonian for the PTMC acts in the model configuration subspace
$\mathrm{Im}P\mathfrak{P}$. Introducing a shorthand notation 
\[
\langle\langle...\rangle\rangle_{l}=\left\langle \Phi_{L}\right\vert ...\left\vert \Phi_{L}\right\rangle 
\]
we easily arrive to

\subsubsection{Ionic contribution to the effective crystal field\label{sub:Ionic-contribution-to}}

It is necessary to average the operators $H_{c}$ and $H_{rr}$ with
the function of the ground state of the $l$-system $\Phi_{L}$. For
$\langle\langle H_{c}\rangle\rangle_{l}$ we get an expression: 
\begin{equation}
\begin{array}{c}
\langle\langle H_{c}\rangle\rangle_{l}=\langle\langle H_{c1}\rangle\rangle_{l}+\langle\langle H_{c2}\rangle\rangle_{l}=\\
=\sum\limits _{\mu}\sum\limits _{i}g_{\mu i}P_{ii}\hat{n}_{\mu}+\sum\limits _{\mu,\nu,\sigma}\sum\limits _{L}V_{\mu\nu}^{L}P_{LL}d_{\mu\sigma}^{+}d_{\nu\sigma}.
\end{array}
\end{equation}
The first term in this expression describes the shifts of the $d$-levels
coming from the interactions of $d$-electrons with electrons on the
$4s$- and $4p$-orbitals of the metal atom. The second term represents
the interaction of $d$-electrons with electrons on the valence orbitals
of the ligands. The sum of the first and the second terms in the expression
eq. () has the form of the operator of the crystal field induced by
the effective charges located on the ligand atoms.

\subsubsection{Covalent contribution to the effective crystal field\label{sub:Covalent-contribution-to}}

Multiplying eq. (\ref{eq:DressedEffectiveHopping}) by $\left\langle \Phi_{L}\right\vert $
from the left and integrating over the variables pertaining to the
$l$-system and taking into account the above averages over the Slater
determinant wave functions: $\delta_{\kappa\lambda}\delta_{\tau\sigma}n_{\kappa}$
and $\delta_{\kappa\lambda}\delta_{\tau\sigma}(1-n_{\kappa})$ in
the first and the second rows of eq. (\ref{eq:DressedEffectiveHopping}),
respectively, we get (the occupation numbers are always unity or zero):
\begin{equation}
\begin{array}{c}
\left\langle \Phi_{L}\right\vert H_{rr}^{(0)}\left\vert \Phi_{L}\right\rangle =\sum\limits _{\mu i\nu j}\sum\limits _{\kappa\tau}\beta_{\mu\kappa}(i)\beta_{\nu\kappa}(j)\times\\
\left\{ -\mathcal{\mathbb{O}}_{-}^{i}\mathcal{\mathbb{O}}_{+}^{j}n_{\kappa}D^{(1+)}(j\kappa)d_{\mu\tau}(i)d_{\nu\tau}^{+}(j)\right.+\\
+\left.\mathcal{\mathbb{O}}_{+}^{i}\mathcal{\mathbb{O}}_{-}^{j}(1-n_{\kappa})D^{(1-)}(j\kappa)d_{\mu\tau}^{+}(i)d_{\nu\tau}(j)\right\} 
\end{array}\label{eq:l-GroundStateEffectiveHopping}
\end{equation}
for the terms with $i\neq j$ and the following:

\begin{equation}
\begin{array}{c}
\left\langle \Phi_{L}\right\vert H_{rr}^{(0)}\left\vert \Phi_{L}\right\rangle =-\sum\limits _{\mu\nu j}\sum\limits _{\kappa\tau}\beta_{\mu\kappa}(j)\beta_{\nu\kappa}(j)\times\\
\left\{ n_{\kappa}D^{(1+)}(j\kappa)d_{\nu\tau}^{+}(j)d_{\mu\tau}(j)-(1-n_{\kappa})D^{(1-)}(j\kappa)d_{\mu\tau}^{+}(j)d_{\nu\tau}(j)\right\} 
\end{array}\label{eq:l-AveragedNeutralResonance}
\end{equation}
for the terms diagonal with respect to $ij$. Clearly only the latter
ones survive the action of the projection operator $\mathfrak{P}$.

\subsubsection{Kinetic exchange and spin polarization}

In this case of an <<inert>> \emph{l}-system the denominator for
the resolvent eq. (\ref{eq:MMCTResolvent}) simplifies to: 
\begin{equation}
\mathfrak{D}_{i\rightarrow j}^{(0)}=\left[\mathfrak{E}-\left(I(i)-A(j)-g_{ij}+E_{R}^{(i\rightarrow j)}\right)\right]^{-1}\label{eq:MagneticResolventDenominatorDiagonal}
\end{equation}
where the subscripts omission as compared to the denominator definition
by eq. (\ref{eq:MagneticResolventDenominator}) indicates its special
form as diagonal with respect to the \emph{l}-system variables. This
quantity is thus the relevant estimate for the effective MMCT energy
as required in works\cite{WeiheGuedel1996,WeiheGuedel1997,WeiheGuedelToftlund,WeihGuedel2000,MossinWeihe}
in the general case. One can easily check that the principal source
of the renormalization of the MMCT energy is the electron-hole interaction
$g_{ij}$ apparently omitted there.\cite{WeiheGuedel1996,WeiheGuedel1997,WeiheGuedelToftlund,WeihGuedel2000,MossinWeihe}
Including the dependence of the ionisation potentials $I(i)$ and
electron affinities $A(j)$ of the \emph{d}-shells on the spins of
the states produced and after assuming the smallness of the intrashell
exchange $K_{dd}$ yields the final formulae\cite{WeiheGuedel1996,WeiheGuedel1997,WeiheGuedelToftlund,WeihGuedel2000}
used for semiquantitative analysis. We do not expand either LMCT,
MLCT, or MMCT energy denominators and use them for calculation in
their precise form.

Further derivation evolves as follows: combining the energy denominator
eq. (\ref{eq:MagneticResolventDenominatorDiagonal}) with the projection
operators relevant to the considered states in the \inputencoding{koi8-r}\foreignlanguage{russian}{$\mathrm{Im}\mathfrak{Q}$
subspace}\inputencoding{latin9} results in the explicit form of the
resolvent: 
\[
\left|\Phi_{L}\right\rangle \left\langle \Phi_{L}\right|\otimes\sum_{ij}\left\{ \mathcal{\mathbb{O}}_{-}^{i}\mathcal{\mathbb{O}}_{+}^{j}\mathfrak{D}_{j\rightarrow i}^{(0)}+\mathcal{\mathbb{O}}_{+}^{i}\mathcal{\mathbb{O}}_{-}^{j}\mathfrak{D}_{i\rightarrow j}^{(0)}\right\} 
\]
The hermitean conjugate of eq. (\ref{eq:DressedEffectiveHoppingLigandDiagonal})
multiplies the above expression on the left and after integrating
over the \emph{l}-system variables yields: 
\begin{alignat}{2}
\left\langle \Phi_{L}\left|\mathfrak{P}H_{rr}^{(0)}\mathfrak{Q}\mathfrak{R}(\mathfrak{E})\mathfrak{Q}H_{rr}^{(0)}\mathfrak{P}\right|\Phi_{L}\right\rangle \approx & \sum\limits _{i\neq j}\sum\limits _{\mu\nu}\sum\limits _{\mu^{\prime}\nu^{\prime}}\nonumber \\
\left\{ \left(\sum_{\kappa}\beta_{\mu^{\prime}\kappa}(i)\beta_{\nu^{\prime}\kappa}(j)D^{(1+)}(j\kappa)n_{\kappa}\right)\right. & \left(\sum_{\lambda}\beta_{\mu\lambda}(i)\beta_{\nu\lambda}(j)D^{(1+)}(j\lambda)n_{\lambda}\right)\mathfrak{D}_{i\rightarrow j}^{(0)} & (a)\nonumber \\
\sum_{\tau\tau^{\prime}}d_{\mu^{\prime}\tau^{\prime}}^{+}(i)d_{\nu^{\prime}\tau^{\prime}}(j)\mathcal{\mathbb{O}}_{-}^{i}\mathcal{\mathbb{O}}_{+}^{j}d_{\nu\tau}^{+}(j)d_{\mu\tau}(i)\nonumber \\
-\left(\sum_{\kappa}\beta_{\mu^{\prime}\kappa}(j)\beta_{\nu^{\prime}\kappa}(i)D^{(1+)}(i\kappa)n_{\kappa}\right) & \left(\sum_{\lambda}\beta_{\mu\lambda}(i)\beta_{\nu\lambda}(j)D^{(1-)}(j\lambda)\left(1-n_{\lambda}\right)\right)\mathfrak{D}_{j\rightarrow i}^{(0)} & (b)\nonumber \\
\sum_{\tau\tau^{\prime}}d_{\mu^{\prime}\tau^{\prime}}^{+}(j)d_{\nu^{\prime}\tau^{\prime}}(i)\mathcal{\mathbb{O}}_{+}^{i}\mathcal{\mathbb{O}}_{-}^{j}d_{\mu\tau}^{+}(i)d_{\nu\tau}(j)\label{eq:Effective-d-Hamiltonian}\\
-\left(\sum_{\kappa}\beta_{\mu^{\prime}\kappa}(j)\beta_{\nu^{\prime}\kappa}(i)D^{(1-)}(i\kappa)\left(1-n_{\kappa}\right)\right) & \left(\sum_{\lambda}\beta_{\mu\lambda}(i)\beta_{\nu\lambda}(j)D^{(1+)}(j\lambda)n_{\lambda}\right)\mathfrak{D}_{i\rightarrow j}^{(0)} & (c)\nonumber \\
\sum_{\tau\tau^{\prime}}d_{\nu^{\prime}\tau^{\prime}}^{+}(i)d_{\mu^{\prime}\tau^{\prime}}(j)\mathcal{\mathbb{O}}_{-}^{i}\mathcal{\mathbb{O}}_{+}^{j}d_{\nu\tau}^{+}(j)d_{\mu\tau}(i)\nonumber \\
+\left(\sum_{\kappa}\beta_{\mu^{\prime}\kappa}(i)\beta_{\nu^{\prime}\kappa}(j)D^{(1-)}(j\kappa)\left(1-n_{\kappa}\right)\right) & \left(\sum_{\lambda}\beta_{\mu\lambda}(i)\beta_{\nu\lambda}(j)D^{(1-)}(j\lambda)\left(1-n_{\lambda}\right)\right)\mathfrak{D}_{j\rightarrow i}^{(0)} & (d)\nonumber \\
\left.\sum_{\tau\tau^{\prime}}d_{\nu^{\prime}\tau^{\prime}}^{+}(j)d_{\mu^{\prime}\tau^{\prime}}(i)\mathcal{\mathbb{O}}_{+}^{i}\mathcal{\mathbb{O}}_{-}^{j}d_{\mu\tau}^{+}(i)d_{\nu\tau}(j)\right\} .\nonumber 
\end{alignat}
The terms of the types (\emph{a}) - (\emph{d}) represent the contributions\cite{WeihGuedel2000}
used there in a simplified form and respectively describe the effects
of: (\emph{a}) transfer of holes between the \emph{d}-shells of the
\emph{i}-th and \emph{j}-th TMIs through the occupied \emph{l}-MOs;
(\emph{b}) and (\emph{c}) <<cyclic>> process consisting of transferring
an electron through the empty \emph{l}-MOs and of the hole through
the occupied \emph{l}-MOs; (\emph{d}) transfer of electrons through
the empty \emph{l}-MOs. While generalizing the configurations\cite{WeihGuedel2000}
these expressions have a sign opposite for the <<cyclic>> processes
(\emph{b}) and (\emph{c}) which in their turn took care about the
phase relations between the MOs in the CN$^{-}$ anions and more sophisticated
bridges. These expressions count for the contribution of every pair
of the \emph{l}-MOs into the transfer process and also take care about
the precised form of the energy (denominators) for the MMCT states.\cite{WeiheGuedel1996,WeiheGuedel1997,WeiheGuedelToftlund,WeihGuedel2000}
Corresponding configurations are depicted in Figs. \ref{fig:TypeAi}-\ref{fig:TypeDi}.

\subsubsection{Spin-correlated double ionization\label{sub:Spin-correlated-double-ionizatio}}

\begin{figure}
\includegraphics{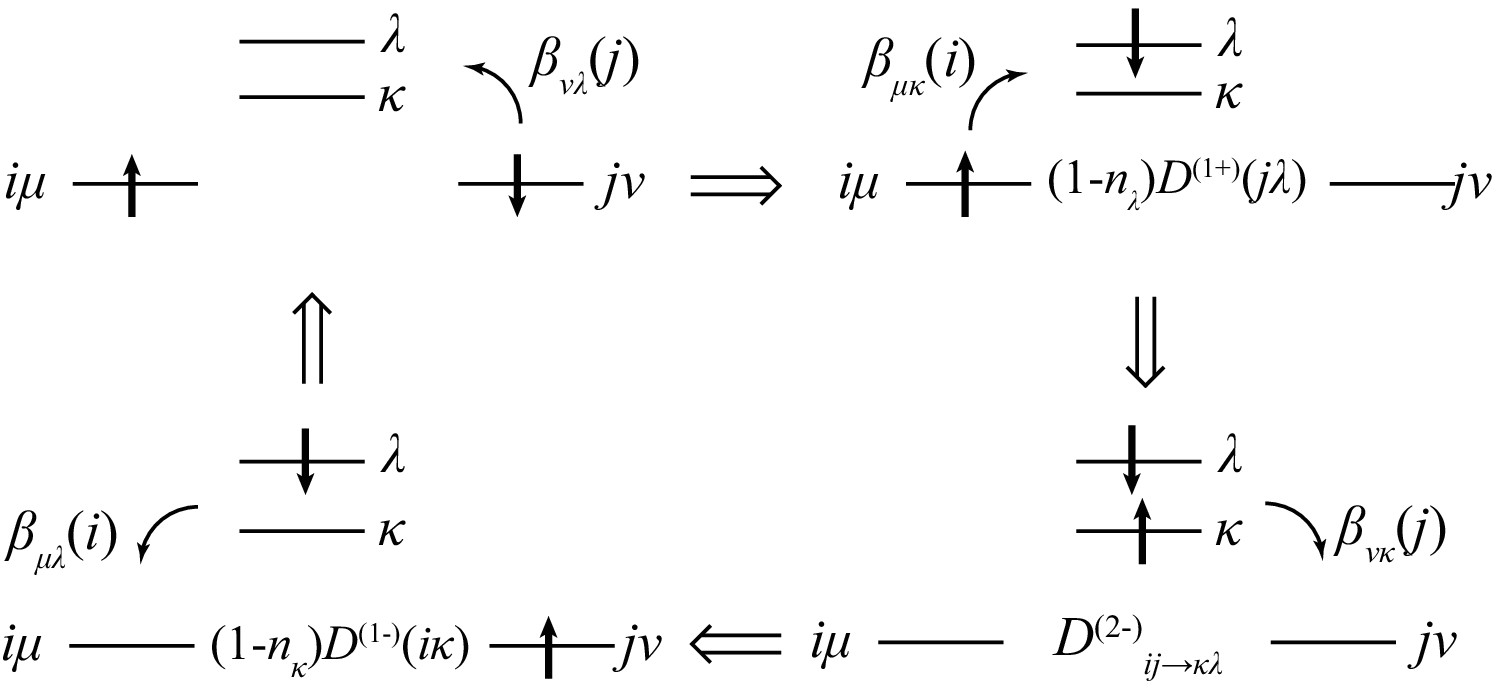}\caption{Sequence of configurations relevant to the matrix elements of type
(f) as contributing to the effective exchange interaction in the case
(i). The configurations in the upper right and lower left corners
belong to the outer subspace $\mathrm{Im}Q^{(1-)}$ - that formed
from the model space configurations by the MLCT states/processes,
that in the lower right corner belongs to the outer subspace $\mathrm{Im}Q^{(2-)}$
- that formed from the model space configurations by the (ML)$^{2}$CT
states/processes. Other notation is as in Fig. \ref{fig:TypeAi}.
\label{fig:TypeFi} }
\end{figure}

Further contributions of the fourth order in $H_{r}$ come from the
ionizing components of the effective operator $H_{rr}$. These are
the averages of the form: $\left\langle \Phi_{L}\left|PH_{rr}^{(2\pm)}QR(E)QH_{rr}^{(2\mp)}P\right|\Phi_{L}\right\rangle $.
Corresponding configurations are depicted in Figs. \ref{fig:TypeEi},
\ref{fig:TypeFi}. The configurations of that type (specifically with
the two electrons extracted from an \emph{l}-MO, but not those with
two electrons added to one of them) have been used,\cite{WeiheGuedelToftlund}
but not systematically. We as usually depart from the action of the
perturbation operator on the ligand ground state $\left|\Phi_{L}\right\rangle $.
For $H_{rr}^{(2+)}$ it reads: 
\[
H_{rr}^{(2+)}\left|\Phi_{L}\right\rangle =\sum\limits _{\mu i\lambda\sigma}\sum\limits _{\nu j\kappa\tau}\beta_{\mu\lambda}(i)\beta_{\nu\kappa}(j)D^{(1+)}(j\kappa)d_{\mu\tau}^{+}(i)d_{\nu\tau}^{+}(j)a_{\lambda\sigma}a_{\kappa\tau}\left|\Phi_{L}\right\rangle .
\]
Multiplying this by the resolvent which in the $\mathrm{Im}Q^{(2+)}$
subspace contains as multipliers the projection operators to doubly
ionized ligand states\\
 $a_{\lambda^{\prime}\sigma^{\prime}}a_{\kappa^{\prime}\tau^{\prime}}\left|\Phi_{L}\right\rangle \left\langle \Phi_{L}\right|a_{\kappa^{\prime}\tau^{\prime}}^{+}a_{\lambda^{\prime}\sigma^{\prime}}^{+}$.
One can expect that the singlet and triplet ionized states contribute
separately. We, however, following\cite{WeiheGuedelToftlund} concentrate
first of all on the states where \emph{each} \emph{l}-MO is doubly
ionized in either sense. That is we take into account only the terms
with $\lambda^{\prime}=\kappa^{\prime}$. In this case $\tau^{\prime}=-\sigma^{\prime}$
holds and inserting the corresponding energy denominator $D^{(2+)}(\lambda^{\prime}\kappa^{\prime}\rightarrow ij)=D^{(2+)}(\kappa^{\prime}\kappa^{\prime}\rightarrow ij)$
which is given by 
\begin{eqnarray}
D^{(2+)}(\lambda\kappa\rightarrow ij) & = & I_{\lambda}-A\left(i\right)-g_{\lambda}\left(i\right)\nonumber \\
 & + & I_{\kappa}-A\left(j\right)-g_{\kappa}\left(j\right)\nonumber \\
 & - & g_{\kappa}\left(i\right)-g_{\lambda}\left(j\right)+g_{\kappa\lambda}+g_{ij}\label{eq:DoublyIonizedEnergies}\\
D^{(2-)}(ij\rightarrow\lambda\kappa) & = & I\left(i\right)-A_{\lambda}-g_{\lambda}\left(i\right)\nonumber \\
 & + & I\left(j\right)-A_{\kappa}-g_{\kappa}\left(j\right)\nonumber \\
 & - & g_{\kappa}\left(i\right)-g_{\lambda}\left(j\right)+g_{\kappa\lambda}+g_{ij}\nonumber 
\end{eqnarray}
we arrive to 
\begin{align}
\sum\limits _{\mu^{\prime}i^{\prime}\lambda^{\prime}\sigma^{\prime}}\sum\limits _{\nu^{\prime}j^{\prime}\kappa^{\prime}\tau^{\prime}}\sum\limits _{\mu i\lambda\sigma}\sum\limits _{\nu j\kappa\tau}\beta_{\mu^{\prime}\lambda^{\prime}}(i^{\prime})\beta_{\nu^{\prime}\kappa^{\prime}}(j^{\prime})\beta_{\mu\lambda}(i)\beta_{\nu\kappa}(j)D^{(2+)}(\lambda^{\prime}\kappa^{\prime}\rightarrow ij)\nonumber \\
D^{(1+)}(j^{\prime}\kappa^{\prime})D^{(1+)}(j\kappa)d_{\mu^{\prime}\tau^{\prime}}(i^{\prime})d_{\nu^{\prime}\tau^{\prime}}(j^{\prime})d_{\mu\tau}^{+}(i)d_{\nu\tau}^{+}(j)\left\langle \Phi_{L}\left|a_{\lambda^{\prime}\sigma^{\prime}}^{+}a_{\kappa^{\prime}\tau^{\prime}}^{+}a_{\lambda\sigma}a_{\kappa\tau}\right|\Phi_{L}\right\rangle \label{eq:CorrelatedIonization}
\end{align}
which contains the averages of the form $\left\langle \Phi_{L}\left|a_{\kappa^{\prime}\tau^{\prime}}^{+}a_{\lambda^{\prime}\sigma^{\prime}}^{+}a_{\lambda\sigma}a_{\kappa\tau}\right|\Phi_{L}\right\rangle $
over the one-determinant\emph{ l}-system ground state which are equal
to\linebreak{}
 $\left(\delta_{\lambda\lambda^{\prime}}\delta_{\kappa\kappa^{\prime}}\delta_{\sigma\sigma^{\prime}}\delta_{\tau\tau^{\prime}}-\delta_{\kappa\lambda^{\prime}}\delta_{\lambda\kappa^{\prime}}\delta_{\tau\sigma^{\prime}}\delta_{\sigma\tau^{\prime}}\right)n_{\kappa}n_{\lambda}$.
Inserting this latter value of the averages and reducing the summation
we arrive to the generalization of the expressions\cite{WeiheGuedelToftlund}
suitable for programming: 
\begin{align}
\sum\limits _{\mu i}\sum\limits _{\nu j}\sum_{\kappa}\left[\beta_{\mu\kappa}(i)\beta_{\nu\kappa}(j)\right]^{2}D^{(2+)}(\kappa\kappa\rightarrow ij)D^{(1+)}(j\kappa)\left[D^{(1+)}(j\kappa)+D^{(1+)}(i\kappa)\right] & (e)\label{eq:CorrelatedIonizationSpecific}
\end{align}
which contributes to the same exchange matrix elements as the Weihe-Güdel
cases (i) and (ii) (see Table \ref{tab:Spin-dependent-factors}).

\subsection{MMCT contributions to the effective spin operator \label{sub:Spin-operator-contributions}}

Finally we are in a position to get the estimates of the contributions
to the effective exchange parameters of the spin Hamiltonian which
is in a way an approximation to the exact eq. (\ref{eq:Effective-d-Hamiltonian}).
We do this in a line with the prescriptions of Ref. \cite{WeiheGuedel1997}
which are based on the fact that if it goes about a binuclear \emph{i.e.}
the simplest possible PTMC (a dimer in the terminology of Refs. \cite{WeiheGuedel1996,WeiheGuedel1997,WeiheGuedelToftlund,WeihGuedel2000})
each of the terms in eqs. (\ref{eq:Effective-d-Hamiltonian}), (\ref{eq:CorrelatedIonization})
yields a contribution to the energy which is proportional to $S(S+1)$
where $S$ is the total spin of the dimer. These contributions when
summed up produce the sought coefficient at the product $\hat{S}_{1}\hat{S}_{2}$
which is the effective exchange parameter. The situation in a PTMC
with more than two TMI's is more complicated; we consider it elsewhere.
For the dimer the derivation evolves as follows: each of the terms
in eq. (\ref{eq:Effective-d-Hamiltonian}) describes a process in
which an electron is taken from one of the \emph{d}-orbitals of one
of the TMI's and put to one of the \emph{d}-orbitals in another TMI.
These orbitals are run over by the subscripts $\mu,\nu$. It is only
possible in high-symmetry systems when an electron taken from one
orbital finally lands in another orbital degenerate with the original
one. However for PTMC (unless it goes about the cuts from highly symmetrical
solids) one cannot count on the representaion with the dimension higher
than two. We, however, assume that the symmetry is low enough so that
no spatially degenerate states can appear. With this precaution we
rewrite the relevant part of the effective operator eq. (\ref{eq:Effective-d-Hamiltonian}).
The expression eq. (\ref{eq:Effective-d-Hamiltonian}) represents
in fact an effective interaction between electrons in the \emph{d}-shells
of the PTMC. The terms off-diagonal with respect to orbital indices
$\mu\mu^{\prime}$ or $\nu\nu^{\prime}$ generate correlated transitions
between the different crystal field states in the respective \emph{d}-shells.
However, the magnetic interactions which are our main goal and thus
the crystal field excitations are not welcome. In fact these terms
lead to various anisotropic contributions, but we for the time being
concentrate on the isotropic ones. In order to avoid excitations other
than magnetic in the ground state manifold one has to set $\left(\mu^{\prime}i\right)=\left(\mu i\right);\left(\mu^{\prime}j\right)=\left(\nu j\right);\left(\nu^{\prime}j\right)=\left(\nu j\right);\left(\nu^{\prime}i\right)=\left(\mu i\right)$
in the above expression and to change $\left(\mu i\right)\leftrightarrow\left(\nu j\right)$
in the terms of the type (\emph{b}) and (\emph{d}). 
\begin{alignat}{2}
\left\langle \Phi_{L}\left|\mathfrak{P}H_{rr}^{(0)}\mathfrak{Q}\mathfrak{R}(\mathfrak{E})\mathfrak{Q}H_{rr}^{(0)}\mathfrak{P}\right|\Phi_{L}\right\rangle \approx & \sum\limits _{i\neq j}\sum\limits _{\mu\nu}\nonumber \\
\left\{ \left(\sum_{\kappa}\beta_{\mu\kappa}(i)\beta_{\nu\kappa}(j)D^{(1+)}(j\kappa)n_{\kappa}\right)\right. & \left(\sum_{\lambda}\beta_{\mu\lambda}(i)\beta_{\nu\lambda}(j)D^{(1+)}(j\lambda)n_{\lambda}\right)\mathfrak{D}_{i\rightarrow j}^{(0)} & (a)\nonumber \\
\sum_{\tau\tau^{\prime}}d_{\mu\tau^{\prime}}^{+}(i)d_{\nu\tau^{\prime}}(j)\mathcal{\mathbb{O}}_{-}^{i}\mathcal{\mathbb{O}}_{+}^{j}d_{\nu\tau}^{+}(j)d_{\mu\tau}(i)\nonumber \\
-\left(\sum_{\kappa}\beta_{\nu\kappa}(j)\beta_{\mu\kappa}(i)D^{(1+)}(i\kappa)n_{\kappa}\right) & \left(\sum_{\lambda}\beta_{\mu\lambda}(i)\beta_{\nu\lambda}(j)D^{(1-)}(j\lambda)\left(1-n_{\lambda}\right)\right)\mathfrak{D}_{i\rightarrow j}^{(0)} & (b)\label{eq:aaa}\\
\sum_{\tau\tau^{\prime}}d_{\nu\tau^{\prime}}^{+}(j)d_{\mu\tau^{\prime}}(i)\mathcal{\mathbb{O}}_{+}^{i}\mathcal{\mathbb{O}}_{-}^{j}d_{\mu\tau}^{+}(i)d_{\nu\tau}(j)\nonumber \\
-\left(\sum_{\kappa}\beta_{\nu\kappa}(j)\beta_{\mu\kappa}(i)D^{(1-)}(i\kappa)\left(1-n_{\kappa}\right)\right) & \left(\sum_{\lambda}\beta_{\mu\lambda}(i)\beta_{\nu\lambda}(j)D^{(1+)}(j\lambda)n_{\lambda}\right)\mathfrak{D}_{i\rightarrow j}^{(0)} & (c)\nonumber \\
\sum_{\tau\tau^{\prime}}d_{\mu\tau^{\prime}}^{+}(i)d_{\nu\tau^{\prime}}(j)\mathcal{\mathbb{O}}_{-}^{i}\mathcal{\mathbb{O}}_{+}^{j}d_{\nu\tau}^{+}(j)d_{\mu\tau}(i)\nonumber \\
+\left(\sum_{\kappa}\beta_{\mu\kappa}(i)\beta_{\nu\kappa}(j)D^{(1-)}(j\kappa)\left(1-n_{\kappa}\right)\right) & \left(\sum_{\lambda}\beta_{\mu\lambda}(i)\beta_{\nu\lambda}(j)D^{(1-)}(j\lambda)\left(1-n_{\lambda}\right)\right)\mathfrak{D}_{i\rightarrow j}^{(0)} & (d)\nonumber \\
\left.\sum_{\tau\tau^{\prime}}d_{\nu\tau^{\prime}}^{+}(j)d_{\mu\tau^{\prime}}(i)\mathcal{\mathbb{O}}_{+}^{i}\mathcal{\mathbb{O}}_{-}^{j}d_{\mu\tau}^{+}(i)d_{\nu\tau}(j)\right\} .\nonumber 
\end{alignat}
Further classification is based on the fact\cite{WeiheGuedel1997,WeiheGuedelToftlund,WeihGuedel2000}
that averages of the terms entering the above expression depend on
the total spin state of PTMC and on the way they are composed of the
states with definite spins of the involved \emph{i}-th and $j$-th
\emph{d}-shells and through this from the occupancy relations between
the involved orbitals. The required calculations based on the work\cite{Damhus}
had been performed.\cite{WeiheGuedel1997} Specifically, each of the
terms in eq. (\ref{eq:aaa}) gives rise to an electron redisribution
process (MMCT) starting and ending in a state of the total spin $S$
with \emph{i}-th and \emph{j}-th \emph{d}-shells occurring in the
states with the spins $S(i)$ and $S(j)$. In the intermediate MMCT
state belonging to the configuration subspace $\mathrm{Im}\mathcal{\mathbb{O}}_{+}^{i}\mathcal{\mathbb{O}}_{-}^{j}\prec\mathrm{Im}\mathfrak{Q}$
the spins of the involved \emph{d}-shells can have only one of four
combinations of the spin values $S(i)\pm\frac{1}{2};S(j)\pm\frac{1}{2}$
which however may realize or not depending on the availability of
electrons to be involved in the specific MMCT process or of free space
in the \emph{d}-shells at hand. The results are given in Table \ref{tab:Spin-dependent-factors}
which is derived\cite{WeiheGuedel1997} under an additional assumption
that the total spin of an individual \emph{d}-shell always goes down
whenever a half-filled orbital is involved in the MMCT process at
this particular shell.

\begin{sidewaystable}
\caption{Spin dependent factors for the numbers $u_{i}$ of unpaired electrons
in the \emph{i}-th \emph{d}-shell.\label{tab:Spin-dependent-factors}}

\begin{tabular}{|c|c|c|c|}
\hline 
Weihe-Güdel  & Change of occupancies & Unpairity  & Spin dependent \tabularnewline
cases & $\left(\mu i\right)\Rightarrow\left(\nu j\right)$ & factors & multipliers\tabularnewline
 & in the MMCT process & $U(u_{i},u_{j})$ & \tabularnewline
\hline 
i  & $\left[1/2\right]\Rightarrow\left[1/2\right]$  & $\frac{2}{u_{i}u_{j}}$ & $+1$ $S(i)\rightarrow S(i)-1/2;S(j)\rightarrow S(j)-1/2$\tabularnewline
\hline 
ii  & $\left[1/2\right]\Rightarrow\left[0\right]$  & $\frac{2}{u_{i}\left(u_{j}+1\right)}$ & $\begin{cases}
\left(+1\right) & S(i)\rightarrow S(i)-1/2;S(j)\rightarrow S(j)-1/2\\
\left(-1\right) & S(i)\rightarrow S(i)-1/2;S(j)\rightarrow S(j)+1/2
\end{cases}$\tabularnewline
\hline 
iii  & $\left[1\right]\Rightarrow\left[1/2\right]$  & $\frac{2}{\left(u_{i}+1\right)u_{j}}$ & $\begin{cases}
\left(+1\right) & S(i)\rightarrow S(i)-1/2;S(j)\rightarrow S(j)-1/2\\
\left(-1\right) & S(i)\rightarrow S(i)+1/2;S(j)\rightarrow S(j)-1/2
\end{cases}$\tabularnewline
\hline 
iv  & $\left[1\right]\Rightarrow\left[0\right]$  & $\frac{2}{\left(u_{i}+1\right)\left(u_{j}+1\right)}$ & $\begin{cases}
\left(+1\right) & S(i)\rightarrow S(i)\pm1/2;S(j)\rightarrow S(j)\pm1/2\\
\left(-1\right) & S(i)\rightarrow S(i)\pm1/2;S(j)\rightarrow S(j)\mp1/2
\end{cases}$\tabularnewline
\hline 
\end{tabular}
\end{sidewaystable}

Otherwise inspecting eq. (\ref{eq:aaa}) shows that for each pair
of the \emph{d}-states $\mu\nu$ located respectively in the \emph{i}-th
and $j$-th TMIs the electron/hole transfer through the empty/occupied
ligand MOs involves formally the same product of the resonance integrals:
\[
\beta_{\nu\kappa}(j)\beta_{\mu\kappa}(i)\beta_{\mu\lambda}(i)\beta_{\nu\lambda}(j)
\]
which depending on the type of the process (a) - (d) is complemented
by a different combination of \emph{l}-MO occupancy numbers and energy
denominators: 
\begin{alignat}{3}
D^{(1+)}(j\kappa) & \mathfrak{D}_{i\rightarrow j}^{(0)} & D^{(1+)}(j\lambda)\times & n_{\kappa}n_{\lambda} & (a)\nonumber \\
-D^{(1+)}(j\kappa) & \mathfrak{D}_{i\rightarrow j}^{(0)} & D^{(1-)}(i\lambda)\times & n_{\kappa}\left(1-n_{\lambda}\right) & (b)\label{eq:aaa-1}\\
-D^{(1-)}(i\kappa) & \mathfrak{D}_{i\rightarrow j}^{(0)} & D^{(1+)}(j\lambda)\times & \left(1-n_{\kappa}\right)n_{\lambda} & (c)\nonumber \\
D^{(1-)}(i\kappa) & \mathfrak{D}_{i\rightarrow j}^{(0)} & D^{(1-)}(i\lambda)\times & \left(1-n_{\lambda}\right)\left(1-n_{\kappa}\right) & (d)\nonumber 
\end{alignat}
Either of the MMCT processes (a) - (d) as classified according to
the ligand MOs involved contributes to the cases (i) - (iv) classified
according to the type of the MMCT process involved. This allows to
establish the correspondence between the general form of the contributions
to the effective exchange parameters and their specific forms and
cases as given in Table \ref{tab:Correspondence-between-terms}. 
\begin{table}
\caption{Correspondence between terms\cite{WeiheGuedelToftlund,WeihGuedel2000}
and eqs. (\ref{eq:CorrelatedIonizationSpecific}), (\ref{eq:aaa}),
(\ref{eq:aaa-1}).\label{tab:Correspondence-between-terms}}

\begin{tabular}{|c|c|}
\hline 
\multirow{2}{*}{Factor in Refs. \cite{WeiheGuedelToftlund,WeihGuedel2000}} & \multirow{2}{*}{Factor in this work }\tabularnewline
 & \tabularnewline
\hline 
$V_{a}$  & $n_{\lambda}\beta_{\mu\lambda}(i)$ or $n_{\kappa}\beta_{\mu\kappa}(i)$ \tabularnewline
\hline 
$V_{a}^{*}$  & $\left(1-n_{\lambda}\right)\beta_{\mu\lambda}(i)$ or $\left(1-n_{\kappa}\right)\beta_{\mu\kappa}(i)$ \tabularnewline
\hline 
$\Delta_{A}$  & $D^{(1+)}(i\kappa)n_{\kappa}$ \tabularnewline
\hline 
$\Delta_{A}^{*}$  & $D^{(1-)}(i\kappa)\left(1-n_{\lambda}\right)$ \tabularnewline
\hline 
$U_{AB}$  & $\mathfrak{D}_{i\rightarrow j}^{(0)}$ \tabularnewline
\hline 
$V_{b}$  & $n_{\lambda}\beta_{\nu\lambda}(j)$ or $n_{\kappa}\beta_{\mu\kappa}(j)$ \tabularnewline
\hline 
$V_{b}^{*}$  & $\left(1-n_{\lambda}\right)\beta_{\nu\lambda}(j)$ or $\left(1-n_{\kappa}\right)\beta_{\mu\kappa}(j)$ \tabularnewline
\hline 
$\Delta_{B}$  & $D^{(1+)}(j\lambda)n_{\lambda}$ \tabularnewline
\hline 
$\Delta_{B}^{*}$  & $D^{(1-)}(j\lambda)\left(1-n_{\lambda}\right)$ \tabularnewline
\hline 
$U_{BA}$  & $\mathfrak{D}_{j\rightarrow i}^{(0)}$ \tabularnewline
\hline 
$\Delta_{AB}$  & $D^{(2+)}(\kappa\kappa\rightarrow ij)$\tabularnewline
\hline 
\end{tabular}
\end{table}

It is remarkable that the overall effect of these contributions \emph{cannot}
be represented as one of the effective hopping of electrons between
the \emph{d}-shells.

\subsection{Correlated double ionization contributions to the effective spin
operator\label{sub:Correlated-double-ionization} }

These terms stem from the averages $\left\langle \Phi_{L}\left|PH_{rr}^{(2+)}Q^{(2+)}R^{(2+)}(E)Q^{(2+)}H_{rr}^{(2+)}P\right|\Phi_{L}\right\rangle $
and are given by eqs. (\ref{eq:CorrelatedIonization}), (\ref{eq:CorrelatedIonizationSpecific}).
The terms within the $(2-)$-segment are processed analogously and
result in 
\[
\sum\limits _{\mu i}\sum\limits _{\nu j}\sum_{\kappa}\left[\beta_{\mu\kappa}(i)\beta_{\nu\kappa}(j)\right]^{2}D^{(2-)}(ij\rightarrow\kappa\kappa)D^{(1-)}(j\kappa)\left[D^{(1-)}(j\kappa)+D^{(1-)}(i\kappa)\right].
\]

\newpage{}

\section*{Figures}


\end{document}